\documentclass[preprint,3p,times,twocolumn,sort&compress]{elsarticle}

\usepackage[T1]{fontenc}
\pdfoutput=1
\journal{Nuclear Inst. and Methods in Physics Research, A}

\usepackage{url}
\usepackage{hyperref}

\usepackage{amsmath}
\usepackage{mathtools}

\usepackage{multirow}
\usepackage{booktabs}
\usepackage{makecell}

\usepackage[format=hang, labelformat=parens]{subcaption}
\usepackage[binary-units]{siunitx}
\begin{document}

\begin{frontmatter}

\title{Using LSTM recurrent neural networks for monitoring the LHC superconducting magnets}

\author[agh_eit]{Maciej Wielgosz}
\ead{wielgosz@agh.edu.pl}

\author[agh_fis,cern]{Andrzej Skocze\'n}
\ead{skoczen@fis.agh.edu.pl}

\author[maribor]{Matej Mertik}
\ead{matej.mertik@um.si}

\address[agh_eit]{Faculty of Computer Science, Electronics and Telecommunications, AGH University of Science and Technology, Krak\'ow, Poland}
\address[agh_fis]{Faculty of Physics and Applied Computer Science, AGH University of Science and Technology, Krak\'ow, Poland}
\address[cern]{The European Organization for Nuclear Research - CERN, CH-1211 Geneva 23 Switzerland}
\address[maribor]{Faculty of Electrical Engineering and Computer Science, University of Maribor, Maribor, Slovenia}

\begin{abstract}
The superconducting LHC magnets are coupled with an electronic monitoring system which records and analyses voltage time series reflecting their performance. A currently used system is based on a range of preprogrammed triggers which launches protection procedures when a misbehavior of the magnets is detected. All the procedures used in the protection equipment were designed and implemented according to known working scenarios of the system and are updated and monitored by human operators.

This paper proposes a novel approach to monitoring and fault protection of the Large Hadron Collider (LHC) superconducting magnets which employs state-of-the-art Deep Learning algorithms. Consequently, the authors of the paper decided to examine the performance of LSTM recurrent neural networks for modeling of voltage time series of the magnets. In order to address this challenging task different network architectures and hyper-parameters were used to achieve the best possible performance of the solution. The regression results were measured in terms of RMSE for different number of future steps and history length taken into account for the prediction. The best result of RMSE=0.00104 was obtained for a network of 128 LSTM cells within the internal layer and 16 steps history buffer.
\end{abstract}

\begin{keyword}
LHC, Recurrent Neural Networks, LSTM, Deep Learning, Modeling
\end{keyword}

\end{frontmatter}

%\linenumbers

\section{Introduction}

The Large Hadron Collider (LHC) is the largest and the most powerful particle collider ever built. It was designed and constructed as a joint effort of the international scientific 
collaboration of the European Organization for Nuclear Research (CERN) \citep{Wright2007, LHCDesRep}. The whole architecture of the LHC is unique and most of its components were custom 
manufactured specifically for this particular application. Consequently, malfunctions and failures of the components usually result in long and costly repairs. This, in turn, affects the 
availability of the particle beams for physics experiments carried out at the LHC. Therefore, maintenance and faults prevention is critical and dedicated solution named Machine Protection 
System (MPS) was created. The MPS system comprises many subsystems, including beam and equipment monitoring, a system to safely stop beam operation and an interlock system 
providing the glue between these systems. The goal is to ensure a safe operation to the accelerator and to maximise a time when particle beams are delivered to interaction points. 

One of the most crucial components of the LHC is a set of superconducting magnets which keep the bunches of protons in a right trajectory inside the vacuum beam pipes in the 
\SI{27}{\kilo\meter} long accelerator tunnel \citep{Wright2007, LHCDesRep}. A voltage on each of the superconducting magnets in the LHC is measured by dedicated digital voltmeter 
\citep{qps} and sent to the central database. The generated stream of the voltage data is used to monitor performance and detect anomalies in the behaviour of superconducting elements. 

One of the most dangerous phenomenon, which can take place at any time in a superconducting electrical circuit, is a quench. It occurs when a part of the superconducting cable becomes 
normally-conducting \citep{cable_stability}. The quench may happen at any time randomly and may occur for many reasons. 
Usually, it is due to a mechanical event inside a superconducting cable or coil, related to the release of stresses generated during production, transportation, and assembly of a magnet. 
Another phenomenon which may lead to a quench is a deposition of energy of particles which escaped from the beam (so called beam losses). When the Quench Protection System (QPS) detects an increased 
resistance, the huge amount of energy stored in the magnet chain is extracted and dumped into a specially designed resistor.

Currently the QPS is the highly dependable system specifically designed for the LHC.  
The instruments of this system perform acquisition of total voltage across superconducting elements (magnet coils, 
bus bars, current leads) and extract resistive component of this voltage. The system \citep{qps, qps_fpga} requires a number of settings. Two the most important settings are:
\begin{itemize}
 \item resistive voltage threshold at which actuators are triggered when the quench event occurs,
 \item discrimination time by which the threshold must be exceeded to recognize the quench event. 
\end{itemize}
The values of these parameters are chosen based on a prior analysis of the magnets and the power supply behaviour. The current approach is very useful 
and has proven its high effectiveness to successfully protect the LHC against severe consequences of the quench events. 

The purpose of this article is to present an approach of modeling the resistive voltage of the LHC superconducting magnets by means of using Recurrent Neural Networks (RNN). The model comprises also such severe phenomena as 
quenches. The existence of so called quench precursors was already observed in voltage signal acquired from superconducting coil \citep{inspiration1, inspiration2}. In the authors' 
opinion, it is likely that a part of the quench causes may gradually unfold in time which ultimately leads the magnet to leave its operating point. If this is the case, they can be modeled 
and predicted using RNN.

Data for training and testing was taken from logging database \citep{timber}.
The low time resolution of this data severely limits the possibility to 
infer a thesis about the effectiveness of quench prediction.
However, the quench data is only used as an example of anomalies occurring during hardware operation - a case study, for which the data was available.
The article demonstrates that even for very low resolution data (one sample for \SI{400}{\milli\second}) the proposed 
neural network structures could model behavior of the magnets.

At so early stage of investigation, there is no chance to answer how fast the network can generate a response. 
The implementation of neural network in FPGA or ASIC is necessary in order to check a timing in comparison to a quench time scale. However, a literature review was conducted in order to estimate the possible system reaction time.

The article contains the following four main contributions:
\begin{itemize}
 \item analysis of LSTM applicability to modeling voltage time series of LHC superconducting magnets,
 \item experimental verification of a range of the LSTM models with real LHC data,
 \item development of a procedure for data extraction and the model training and testing,
 \item development of a custom designed prototype application of LSTM-based model for conducting experiments.
\end{itemize}

The rest of the article is structured as follows. 
Section~\ref{section:lhc} provides the background and related work of the LHC, superconducting magnets and quench protection. 
Section~\ref{section:rnn} contains theory and mathematical formalism for recurrent neural networks. Section~\ref{section:visual} presents an idea of a visualization environment for 
the results of the experiments. 
Section~\ref{section:experiments} describes the architecture of the custom designed system used for the experiments as well as for data acquisition and provides the results of 
the experiments. Section~\ref{section:discussion} provides a discussion about possible uses of the proposed solution, as well as the performance of comparable systems. 
Finally, the conclusions of our research are presented in Section~\ref{section:conclusions}.

\section{Large Hadron Collider}
\label{section:lhc}
Large Hadron Collider (LHC) is currently the most powerful scientific instrument ever built. The main objective of this huge enterprise is a pure desire for exploring the deepest structure of matter. The project was launched in 1994 and it is managed by the European Organization for Nuclear Research (CERN). After overcoming many technical challenges, the LHC started the operation in 2010 \citep{LHCDesRep}. Currently the second run of the LHC is in progress.
 
One of the ideas used in high energy physics experiments is essentially based on a concept of colliding two high energy particles which travel in opposite directions. This allows to look deeply into the structure of the matter which constitutes our universe. The particles used in case of the LHC are protons or lead ions. The products of protons collision are carefully analysed by huge systems of particle detectors. There are four main independent detection systems built at the LHC, namely: ATLAS, CMS, LHCb and ALICE. 
    
The main goal of the LHC is confirmation or refutation of the theories in the field of elementary particle physics. One of the most crucial questions which LHC was intended to address is related to the existence of the postulated Higgs boson which was ultimately discovered in the course of the experiments conducted independently by two collaborations at CERN in 2012 \citep{Higgs_ATLAS, Higgs_CMS}.
   
For the most part, the LHC is located in the tunnel lying between \SI{45}{\meter} and \SI{170}{\meter} below the earth's surface near Geneva lake. The tunnel is of a circular shape with the circumference of approx. \SI{27}{\kilo\meter}. 
There are many superconducting magnets located in the tunnel which provide a magnetic field necessary to lead the proton beams around a circular trajectory. The LHC tunnel is divided into eight different sectors. 
The particles are injected into the LHC with energy of \SI{450}{\giga\electronvolt}. They are prepared within the smaller accelerator (called SPS) and injected into the LHC in bunches. 
A single bunch contains nominally \num{1.15d11} protons. The operation of gradual delivery of proton bunches to the LHC is denoted as ``filling the machine''. It takes \num{2808} bunches altogether to fill up the LHC. The time between bunches is \SI{25}{\nano\second}. 

It is worth noting that all the bunches travelling along the LHC circle are accelerated in one dedicated place. The remaining sections of the circle guide particles to the accelerating cavities and during each revolution the energy of particles is raised. In order to maintain a stable trajectory of the particles the uniform dipole magnetic field has to be raised synchronously with the raising particle energy. This in turn results in a ramp up of a current in the superconducting dipoles. The described process, denoted as ``ramping up the machine'', allows to achieve a particle energy of \SI{7}{\tera\electronvolt} after multiple iterations. Tab.~\ref{tab:LHClevels} shows initial and final levels of proton's energy, magnetic field and current supply. 

When desired energy is achieved, the beams collide at four points around the circle which are surrounded by the four detection systems. There is a huge amount of data produced by the detectors since every \SI{25}{\nano\second} two bunches collide giving a number of individual proton-proton collisions. The tracks of particles produced in each individual collision are recorded by detection system. The data gathered by the system is processed by a reconstruction algorithm.

\begin{table}
\caption{The nominal conditions in the main dipole circuits of the LHC at the beginning and at the end of ramping up \citep{LHCDesRep}.}
\label{tab:LHClevels}
\centering
\begin{tabular}{lccc}
\toprule
Parameter & Injection & Collision & Unit \\
\midrule
Proton energy & \num{0.450} & \num{7} & \si{\tera\electronvolt} \\
Magnetic field & \num{0.535} & \num{8.33} & \si{\tesla} \\
Supply current & \num{763} & \num{11850} & \si{\ampere} \\
\bottomrule
\end{tabular}
\end{table}
   
\subsection{Superconducting magnets}
The superconducting magnets are the critical components of the LHC which store huge amount of magnetic energy. 
This imposes a series of challenges related to powering the whole system. The cables used to wind the magnet coils and to 
deliver a current (bus bars, current leads) to the coils must conduct the current at the level of $\approx \SI{12000}{\ampere}$ in the magnetic field of $\approx \SI{8.5}{\tesla}$ (Tab.~\ref{tab:LHClevels}). 
Consequently, the designers decided to take advantage of superconducting materials which meet all the electrical and magnetic requirements of the LHC. The instruments built with superconductors are also small enough to fit in the tunnel. 

The superconducting cables are not cryostable and therefore a random and local temperature change can lead to a sudden transition to a normal conduction state \citep{cable_stability}. This phenomenon is known as quench. 
During assembly and operation, a superconducting coil is always subjected to stresses resulting from pre-loading at assembly, from differential thermal contractions at cool-down and from the electromagnetic forces during its regular operation. 
The release of the mechanical energy happens locally, through micro-slips constrained by friction,  vibration, or local cracking. The amount of energy generated within the process can be enough to elevate temperature locally above a critical value. 
Consequently, the resistive place in the cable generates enough heat to damage the cable.

Beam losses are a very important problem for superconducting accelerators. Protons which escape from a bunch in the direction perpendicular to the beam hit the wall of the vacuum beam pipe. 
Cascades of particles are produced and a radiated energy is deposited in the surrounding materials, in particular in the superconducting windings. 
This energy can locally heat the coil above critical temperature, causing a quench \citep{beam_losses}. 

Quenches may occur in various circumstances but some of the most common ones take place during a so-called magnet training. 
At the first powering during ramping up a current, a magnet losses superconducting state long before reaching the expected critical current. 
At the next attempt of powering, the current that could be reached before quench is higher. The process continues over all the next attempts, and the maximum current that could be reached increases quench after quench, slowly approaching a plateau.

A circular particle accelerator requires a dipole magnetic field to maintain the particle beam within its trajectory. Furthermore, several other kinds of magnets are required for shaping and guiding the beam. In the case of the LHC, most of them are superconducting magnets supplied with constant current by means of power converters. A summary of superconducting circuits is presented in Tab.~\ref{tab:LHCcircuits}.

\begin{table}[]
\centering
\caption{The general overview of the circuits powering the superconducting magnets of the LHC \citep{LHC_layout, LHC_quench_db}. The number of quenches is reported on 10 October 2016.}
\label{tab:LHCcircuits}
\begin{tabular}{lccc}
\toprule
\makecell[l]{LHC\\Circuit} & \makecell{No of\\circuits} & \makecell{No of magnets\\ in one circuit} & \makecell{No of\\quenches} \\
\midrule
RB & \num{8} & \num{154} & \num{1270} \\
RQ  & \num{16} & \num{47} & \num{64} \\
IT & \num{8} & \num{4} & \num{18} \\
IPQ & \num{78} & \num{2} & \num{323} \\
IPD & \num{16} & \num{1} & \num{53} \\
\midrule
\SI{600}{A} EE & \num{202} & $m$ & \multirow{3}{*}{\num{425}} \\
\SI{600}{A} EEc & \num{136} & \num{1} or \num{2} & \\
\SI{600}{A} & \num{72} & \num{1} & \\
\midrule
$\num{80}\div\SI{120}{A}$ & \num{284} & \num{1} & \num{116} \\
\SI{60}{A} & \num{752} & \num{1} & \num{44} \\ 
\bottomrule
\end{tabular}
\flushleft
\footnotesize{RB - Main Dipole; RQ - Main Quadrupole; IT - Inner Triplet; IPQ - Individually Powered Quadrupole; IPD - Individually Powered Dipole; EE - Energy Extraction; EEc - Energy Extraction by crowbar; $m$ - number of magnets in circuits is not constant in this class of circuits.} 
\end{table}
   
\subsection{Quench protection}
\label{subsection:qps}
A need for a system of an active magnet protection originates from the nature of the superconducting cables used to build the magnets. Most of the high-current superconducting magnets used in the LHC are not self-protected and would be damaged or destroyed if they were not protected during the quench. Therefore, the quench protection system was introduced \citep{qps, qps_fpga}.
The LHC Machine Protection System (MPS) comprises many subsystems. One of the subsystems is a Quench Protection System (QPS). 
This system consists of a Quench Detection System (QDS) and actuators which are activated once a quench is detected. 

\begin{figure}
\centering
\includegraphics[width=1\hsize]{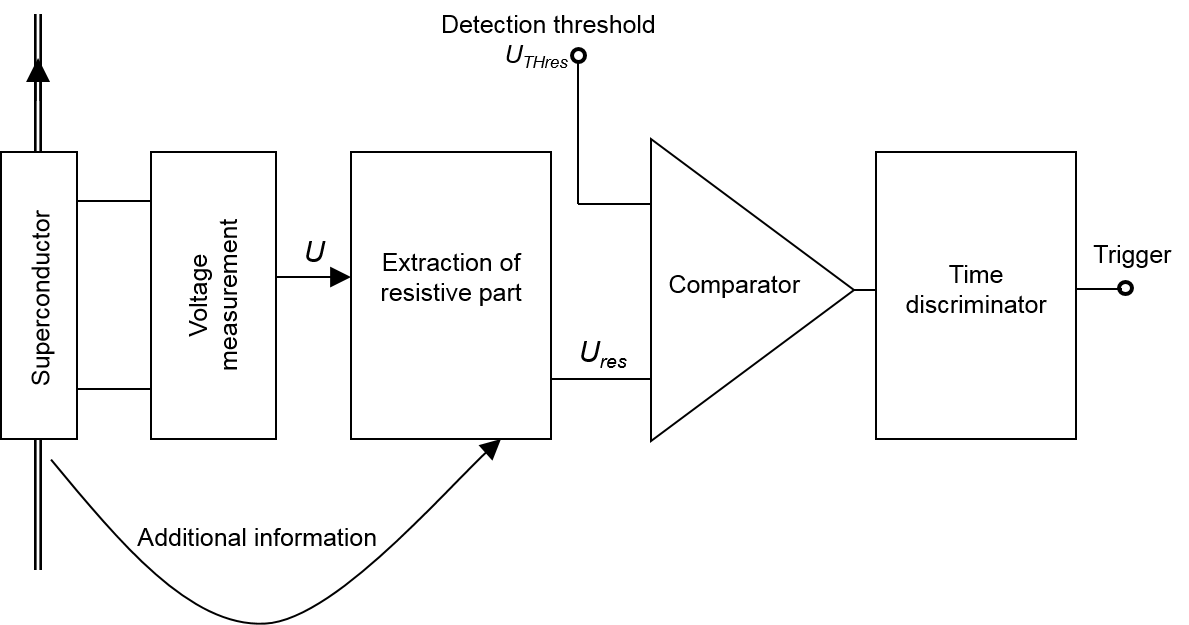}
\caption{The general architecture of the quench detector.}
\label{fig:qdet_principle}
\end{figure}

A superconducting magnet has zero resistance and a relatively large inductance which is equal $\approx \SI{100}{\milli\henry}$ in the case of main LHC dipole.  When a constant current flows through the magnet, the total voltage across it, is zero. When the magnet loses its superconducting state (quench) the resistance becomes non-zero, hence, a voltage develops over the resistive part. 
This voltage is used to detect the quench. However, during normal operation (ramp up or down, fast power abort) a current change in the magnet generates an inductive voltage which might be well above the resistive voltage detection threshold. 
Therefore, the inductive voltage must be compensated in order to prevent the QDS from spurious triggering. Consequently, the most important part of the quench detector is an electronic module for extracting the resistive part of the total voltage.
It is shown in Fig.~\ref{fig:qdet_principle}. 

A quench detector is an electronic device with the following functions (Fig.~\ref{fig:qdet_principle}): 
\begin{itemize}
\item Monitoring of the voltage of superconducting elements, 
\item Extraction of the resistive part of the voltage $U_{res}$, 
\item Generation of trigger signals in case the resistive voltage exceeds the threshold. 
\end{itemize} 
The triggers are transmitted to other protection devices via current loops to initiate a safe shutdown of the electric circuits supplying the magnets.

The method for compensation of the inductive voltage is simple in the case of a differential magnet circuit where two very similar inductances are connected in series in one circuit. 
However, in some LHC corrector magnet circuits, there are no reference elements available, hence the compensation of the inductive voltage by simple subtraction cannot be implemented. 
In such a case, Kirchhoff's voltage law for the circuit must be solved. To satisfy timing requirements, the solution must be performed numerically (online) by means of using digital logic. 

\begin{figure}
\centering
\includegraphics[width=1\hsize]{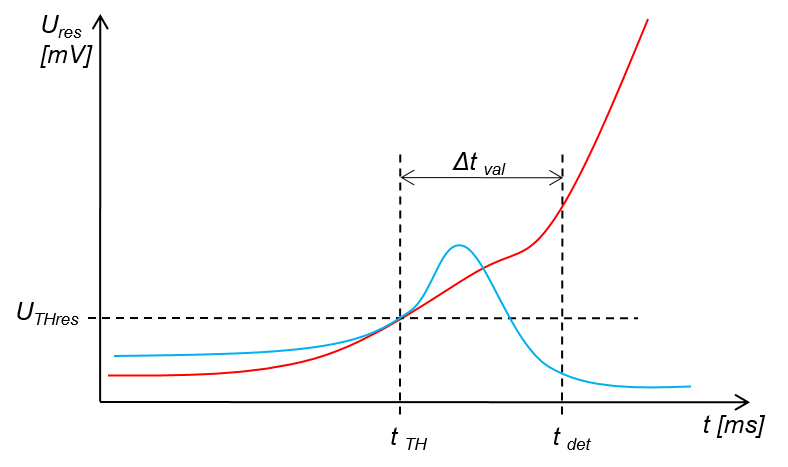}
\caption{The principle of a quench validation by means of using a time discriminator.}
\label{fig:waveform_principle}
\end{figure}

Next, a quench candidate is validated as a real quench or noise. This is carried out by means of a time discriminator shown in Fig.~\ref{fig:qdet_principle}. 
The voltage resistive component $U_{res}$ must be higher than a threshold for the time interval longer than a validation time $\Delta t_{val}$ in order to be classified as a quench. This condition is depicted in Fig.~\ref{fig:waveform_principle}. 

The trigger signal has two important functions: the release of energy to quench heaters and an opening for an interlock loop. The goal of the quench heater is an acceleration of the propagation of the quench along a cable. It prevents local overheating (or even melting) of the quenching cable. The opening of the interlock loop is a method for an immediate transferring of the request for the termination of an operation of other LHC components.
 
Voltage time series measured and extracted by the QPS system are sent over to two different storage systems. The system called the CERN Accelerator Logging Service (CALS) contains low resolution data \citep{timber}. 
The second system, called POST\_MORTEM, is dedicated to store data delivered by any equipment in the LHC whenever a trigger occurs \citep{pmdata}.  

\section{Recurrent neural networks}
\label{section:rnn}
Recent years have witnessed a huge growth of deep learning applications and algorithms. They are powerful learning models, which achieve great successes in many fields and win multiple competitions \citep{krizhevsky}.
The neural nets are capable of capturing latent content of the modeled objects in large hierarchies \citep{leCun_deep_2015, bookAGraves, Schmidhuber_Deep_2014, deng_deep_2014}. 
Two main branches of the neural networks are feed-forward and recurrent models. 
The members of the first one, of which Convolutional Neural Networks (CNN) is now the most prominent example, are usually used for processing data belonging to a spatial domain, where data occurrence in time is not important and not taken into account \citep{leCun_deep_2015, Farabet_Learning_2013, LeCun_deep_learning_2015}. Opposed to that there are algorithms working in temporal domain, in which the information about the order of data is critical. 

Since magnets behavior modeling involves temporal dependencies of the examined signals we decided to focus on the neural network models that are capable of sequence processing, namely RNN, Long Short-Term Memory (LSTM) and Gated Recurrent Unit (GRU) \citep{bookAGraves, Morton_2016_Analysis, Pouladi_2015_Recurrent, Chen_2016_Efficient}.

Currently, the LSTM is considered to be the best model \citep{Hochreiter_Long_1997} and it is also the most often used in applications. Therefore, we have decided to use it in our experiments.

Unlike traditional models that are mostly based on hand-crafted features deep learning neural networks can operate directly on raw data. This makes them especially useful in applications where extracting features is very hard and even sometimes impossible. It turns out that there are many fields of applications where no experts exist who can handle feature extraction or the area is simply uncharted and we do not know whether the data contains latent patterns worth exploring \citep{Chang_2016_Application, Cai_2013_Deep, Toth_2013_Phone}. 

Foundations of the most neural network architectures currently used were laid down between 1950 and 1990. For almost the last two decades researches were not able to take full advantage of these powerful models. But the whole machine learning landscape changed in early 2010, when deep learning algorithms started to achieve state-of-the-art results in a wide range of learning tasks. The breakthrough was brought about by several factors, among which computing power, huge amount of widely available data and affordable storage are considered to be the critical ones. It is worth noting that in the presence of large amount of data, the conventional linear models tend to under-fit or under-utilize computing resources. 

\begin{figure}
\centering
\includegraphics[width=0.3\hsize]{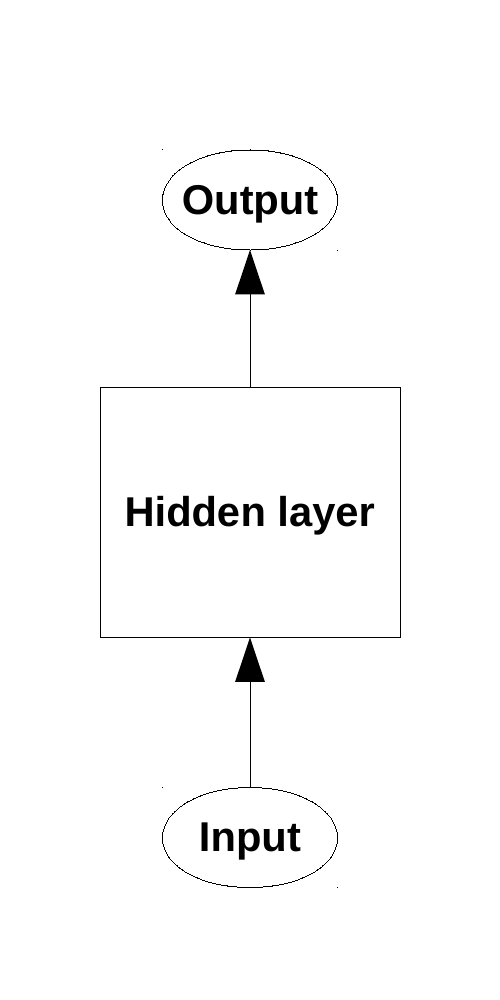}
\caption{The architecture of standard feed-forward neural network.}
\label{fig:feedforward_overall_diagram}
\end{figure}

CNNs and feed-forward networks rely on the assumption of the independence of data within training and testing set as presented in Fig.~\ref{fig:feedforward_overall_diagram}. This means that after each training item is presented to the model, the current state of the network is lost i.e. temporal information is not taken into account in training a model. 

In the case of independent data, it is not an issue. But for data which contain crucial time or space relationships, it may lead to the loss of the majority of the information which is located in between steps. Additionally, feed-forward models expect a fixed length of training vectors which is not always the case, especially when dealing with time domain data.

\begin{figure*}
\centering
\includegraphics[width=1\hsize]{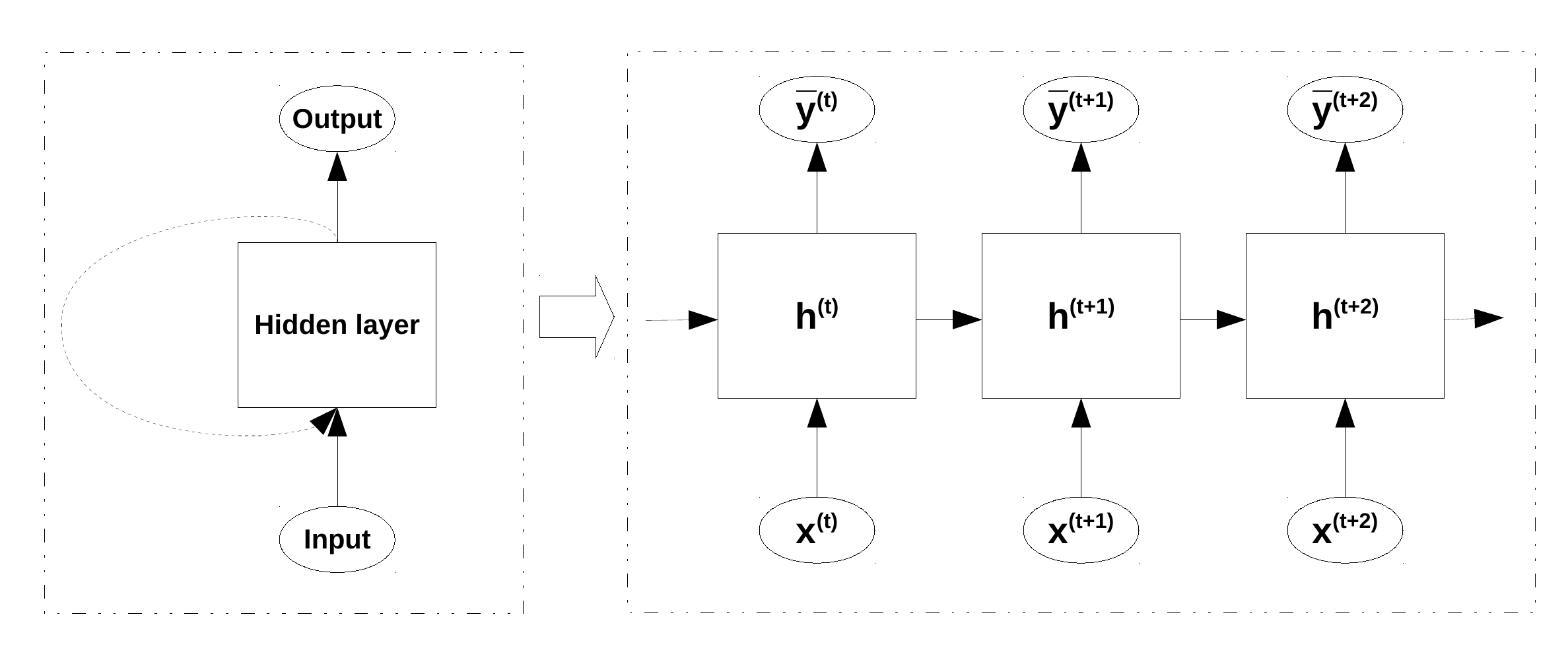}
\caption{The general overview of recurrent neural networks.}
\label{fig:rnn_overall_diagram}
\end{figure*}

\begin{figure}
\centering
\includegraphics[width=.35\hsize]{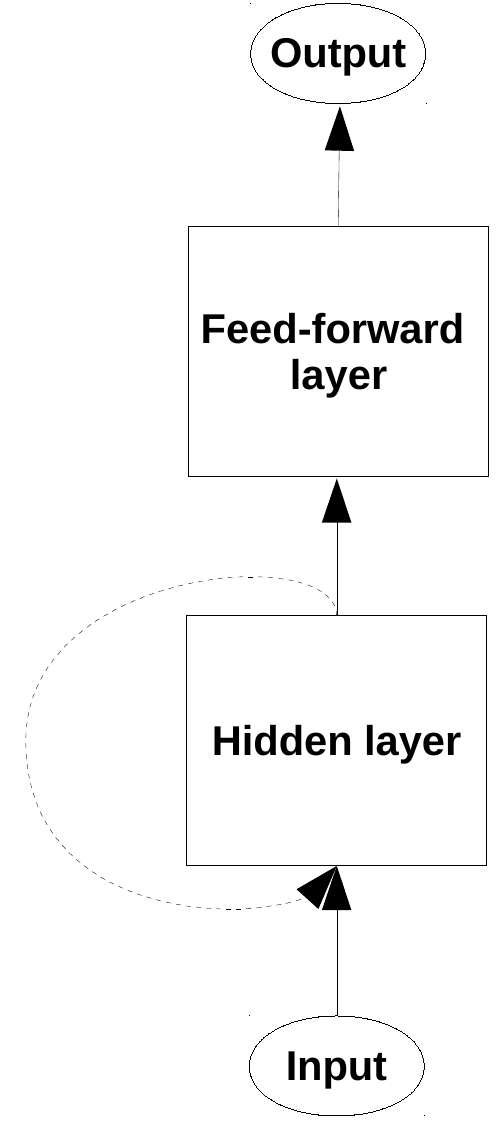}
\caption{The mapping of outputs of a recurrent neural network.}
\label{fig:ffn_over_rnn_diagram}
\end{figure}

\begin{figure}
\centering
\includegraphics[width=0.9\hsize]{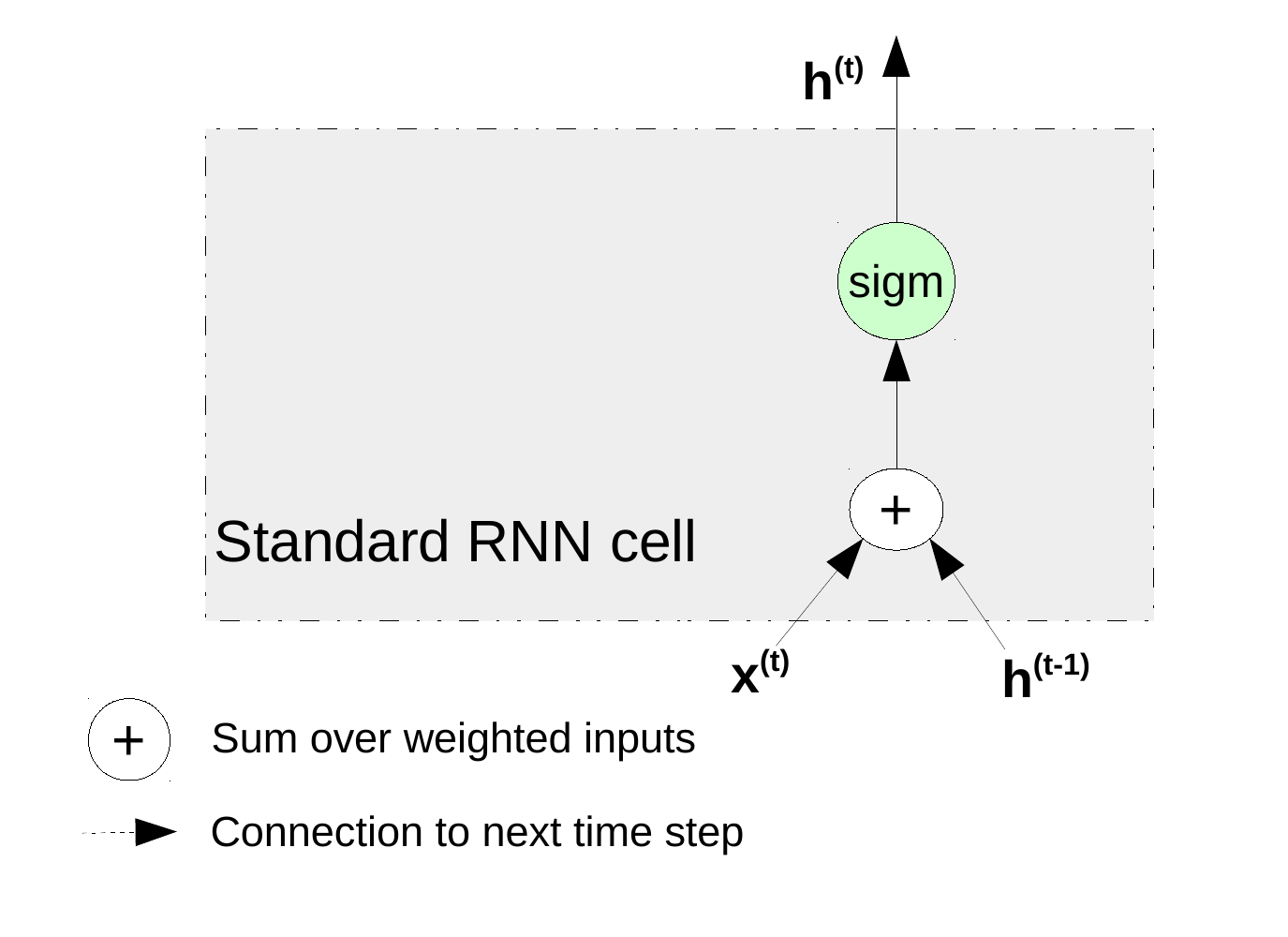}
\caption{A cell of a standard recurrent neural network.}
\label{fig:rnn_block_architecture}
\end{figure}

Recurrent neural networks (RNNs) are models with the ability to process sequential data one element at a time. Thus they can simultaneously model sequential and time dependencies on multiple scales. Unfortunately, a range of practical applications of standard RNN architectures is quite limited. This is caused by the influence of a given input on hidden and output layers during the training of the network. It either decays or blows up exponentially as it moves across recurrent connections. This effect is described as the \textit{vanishing or exploding gradient problem} \citep{bookAGraves}. 
There had been many unsuccessful attempts to address this problem before LSTM was eventually introduced by Hochreiter and Schmidhuber \citep{Hochreiter_Long_1997} which ultimately solved it.

Recurrent neural networks may be visualized as looped-back architectures of interconnected neurons. This was presented in Fig.~\ref{fig:rnn_overall_diagram}. Originally RNNs were meant to be used with single variable signals but they have also been adapted for multiple stream inputs \citep{bookAGraves}.

It is a common practice to use feed-forward network on top of recurrent layers together in order to map outputs from RNN or LSTM to the result space as presented in Fig.~\ref{fig:ffn_over_rnn_diagram}. 

\subsection{RNN}

Architecture of standard neural networks is presented in Fig.~\ref{fig:rnn_block_architecture}. The nodes of the network receive input from the current data point $x^{(t)}$ as well as the hidden state values of the hidden layer in the previous state $h^{(t-1)}$. Thus, inputs at time $t$ have impact on the outputs of the network to come in the future by the recurrent connections. 

There are two fundamental equations (\ref{eq:rnn_ht}) and (\ref{eq:rnn_yt}), which characterize computations of a recurrent neural network as presented in Fig.~\ref{fig:rnn_block_architecture}.

\begin{equation}
h{(t)} = Q(W_{(hx)}x^{(t)} + W_{(hh)}h^{(t-1)} + b_h),
 \label{eq:rnn_ht}
\end{equation}

\begin{equation}
\hat{y} ^{(t)} = \sigma(W_{(yh)}h^{(t)} + b_y)
 \label{eq:rnn_yt}
\end{equation}

\noindent
where: $Q$ is an activation function. $W_{(hx)}$, $W_{(yh)}$ and $W_{(hh)}$ are weights matrices of input-hidden layer, hidden-output layer and recurrent connections respectively. $b_h$ and $b_y$ are vectors of biases. 

Standard neural networks are trained across multiple time steps using the algorithm called \textit{backpropagation through time} \citep{Zachary_2015_Critical}.

\subsection{LSTM}

\begin{figure}
\centering
\includegraphics[width=1\hsize]{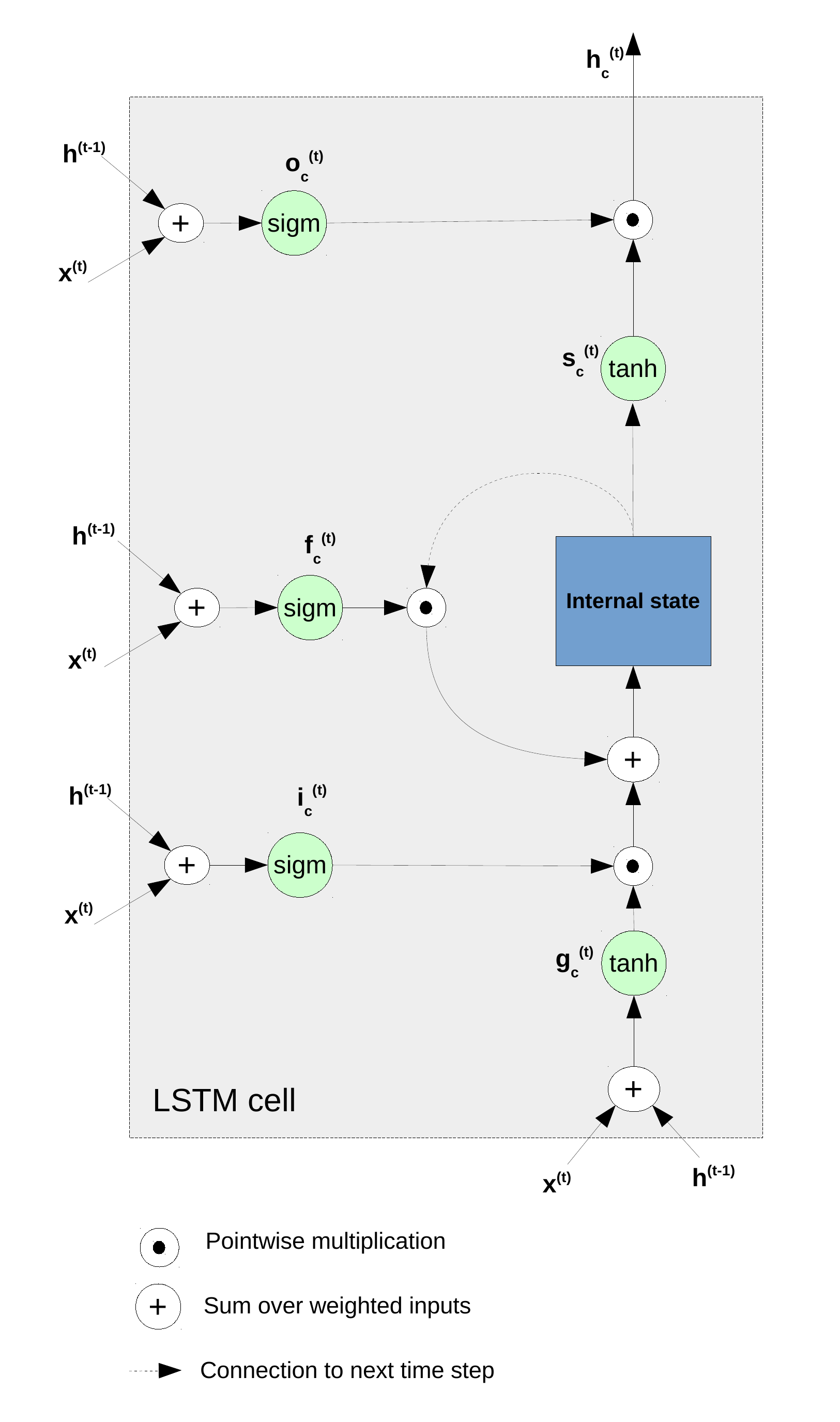}
\caption{An architecture of the LSTM cell}
\label{fig:lstm_block_architecture}
\end{figure}

In practical applications, the Long Short-Term Memory (LSTM) model has shown extraordinary ability to learn long-range dependencies as compared to standard RNNs. Therefore, most of state-of-the-art applications use the LSTM model \citep{Zachary_2015_Critical}. 

The LSTM internal structure is based on a set of connected cells. The structure of a cell is presented in Fig.~\ref{fig:lstm_block_architecture}, it contains feedback connection storing the temporal state of the cell. Additionally, the LSTM cell contains three gates and two nodes which serve as an interface for information propagation within the network. 

There are three different gates in each LSTM cell:
\begin{itemize}
 \item \textit{input} gate $i_c^{(t)}$ which controls input activations into the memory element,
 \item \textit{output} gate $o_c^{(t)}$ controls cell outflow of activations into the rest of the network,
 \item \textit{forget} gate $f_c^{(t)}$ scales the internal state of the cell before summing it with the input through the self-recurrent connection of the cell. This enables gradual forgetting in the cell memory.
\end{itemize}
\noindent
In addition, the LSTM cell also comprises an input node $g_c^{(t)}$ and an internal state node $s_c^{(t)}$.

Modern LSTM architectures may also contain \textit{peephole connections} \citep{Greff_2015_LSTM}. Since they are not used in the experiment, they were neither depicted in Fig.~\ref{fig:lstm_block_architecture} nor addressed in this description.

The output of a set of LSTM cells is calculated according to the following set of vector equations:

\begin{equation}
g^{(t)} = \phi(W_{gx}x^{(t)} + W_{gh}h^{(t-1)} + b_g),
 \label{eq:lstm_input_node}
\end{equation}

\begin{equation}
i^{(t)} = \sigma(W_{ix}x^{(t)} + W_{ih}h^{(t-1)} + b_i),
 \label{eq:lstm_input_gate}
\end{equation}

\begin{equation}
f^{(t)} = \sigma(W_{fx}x^{(t)} + W_{fh}h^{(t-1)} + b_f),
 \label{eq:lstm_forget_gate}
\end{equation}

\begin{equation}
o^{(t)} = \sigma(W_{ox}x^{(t)} + W_{oh}h^{(t-1)} + b_o)
 \label{eq:lstm_output_gate}
\end{equation}

\begin{equation}
s^{(t)} = g^{(t)} \odot i^{(t)} + s^{(t-1)} \odot f^{(t)},
 \label{eq:lstm_internal_state}
\end{equation}

\begin{equation}
h^{(t)} = \phi (s^{(t)}) \odot o^{(t)}.
 \label{eq:lstm_output_node}
\end{equation}

While examining equations (\ref{eq:lstm_input_node}) -- (\ref{eq:lstm_output_node}), it may be noticed that instances for a current and previous time step are used for the value of the output vector of hidden layer  $h$ as well as for the internal state vector $s$. Consequently, $h^{(t)}$ denotes a value of an output vector at the current time step, where as $h^{(t-1)}$ refers to the previous step. It is also worth noting that the equations contain vector notation which means that they address the whole set of LSTM cells. In order to address a single cell a subscript $c$ is used as it is presented in Fig.~\ref{fig:lstm_block_architecture}, where for instance $h_c^{(t)}$ refers to a scalar value of an output of this particular cell.

The LSTM network learns when to let an activation into the internal states of its cells and when to let an activation of the outputs. In this gating mechanism all the gates are considered as separate components of the LSTM cell with their own learning capability. This means that the cells adapt during training process to preserve a proper information flow throughout the network as separate units. Therefore, when the gates are closed, the internal cell state is not affected. In order to make this possible a hard sigmoid function $\sigma$ was used, which can output \num{0} and \num{1} as given by equation (\ref{eq:hard_sigma}). As a result the gates can be fully opened or fully closed.

\begin{equation}
\begin{split}
\sigma (x) =
\begin{cases}  
0 \text{~if~}  x \leq t_{l}, \\ 
ax + b  \text{~if~} x \in (t_{l}, t_{h}), \\
1 \text{~if~}  x \geq t_{h}. \\ 
\end{cases}
\label{eq:hard_sigma}
\end{split}
\end{equation}

In terms of the backward pass, so-called constant error carousel enables the gradient to propagate back through many time steps  \citep{Hochreiter_Long_1997, Zachary_2015_Critical}. 

\section{Visualization framework}
\label{section:visual}

The model is intended to be integrated within visualisation environment for the experiments. Python framework based on Django (storing and managing experiments setup data) \citep{Django} and Bokeh (interactive Python library for visualisation of data) \citep{Bokeh} will be used for the development of web application for quench prediction. Described LSTM model will be integrated to building blocks of an ELQA data analysis framework \citep{Barnard_usabilityof} developed at Machine Protection and Electrical Integrity group (TE-MPE) in order to prototype a web based quench prediction application for use at CERN.  

In the ELQA framework, an access to the data is addressed with the Object-Relational Mapping. The Django framework handles this mapping and provides full functionality of the Structure Query Language (SQL). The architecture of the Django is organized with three layers as follows:
\begin{itemize} 
\item the bottom layer which is a database, followed by 
\item an access library that is responsible for a communication between Python and the database by means of SQL statements and 
\item a specific Django database back-end.  
\end{itemize}
\noindent

As the access library to CERN Oracle from Django a Python library \verb+cx_oracle+ \citep{cxOracle} is used. For results capturing and maintaining appropriate data model is defined, which can be created by means of the tool called \verb+inspectdb+ available inside Django. Information necessary for this process is taken from database tables, however relationship between the tables should be separately defined. Dashboard for an application will be designed with widgets and plots available within ELQA framework and Bokeh library.

\section{Experiments}
\label{section:experiments}
This section presents the results of the experiments which were conducted in order to validate performance of the LSTM network in an voltage time series modeling task. All the experiments required several steps of preparation which were mostly related to a data preprocessing. 

\subsection{Setup description}
\label{subsection:experiments_setup}

Data acquired from all the CERN accelerators are kept in the global database called Logging Service (LS) \citep{timber}. Despite the very low time resolution (one sample for \SI{400}{\milli\second}) of this data, it is possible to examine the feasibility of using LSTM for modeling the behavior of the magnets. A generic Java GUI called TIMBER \citep{LHC_logging_db} and a dedicated Python wrapper \citep{pytimber} are provided as tools to visualize and extract logged data.

The logging database stores a record of many years of the magnets activity. This is a huge amount of data with relatively few quench events. Since part of the planned research focuses on quench prediction and/or detection it was important to include such a data in constructed datasets. It is worth noting that one day-long record of single voltage time series for a single magnet occupies roughly \SI{100}{\mega\byte}. There are several voltage time series associated with a single magnet \citep{LHC_logging_db} but ultimately authors decided to use $U_{res}$ in the experiments. The origin and the meaning of the resistive voltage $U_{res}$ were discussed in the subsection \ref{subsection:qps}.

\begin{figure*}
\centering
\begin{subfigure}{.4\textwidth}
  \centering
  \includegraphics[width=.9\linewidth]{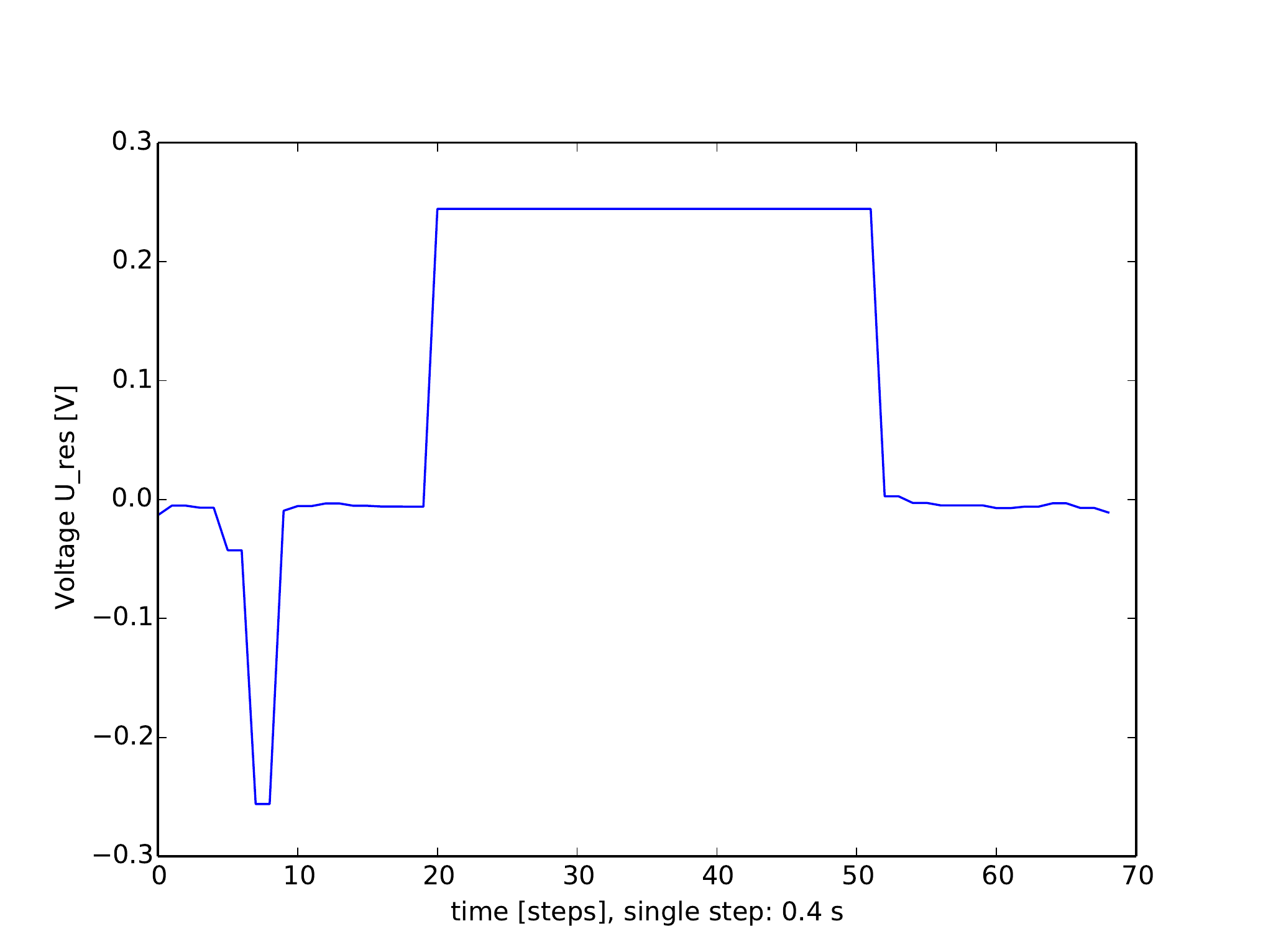}
\end{subfigure}%
\begin{subfigure}{.4\textwidth}
  \centering
  \includegraphics[width=.9\linewidth]{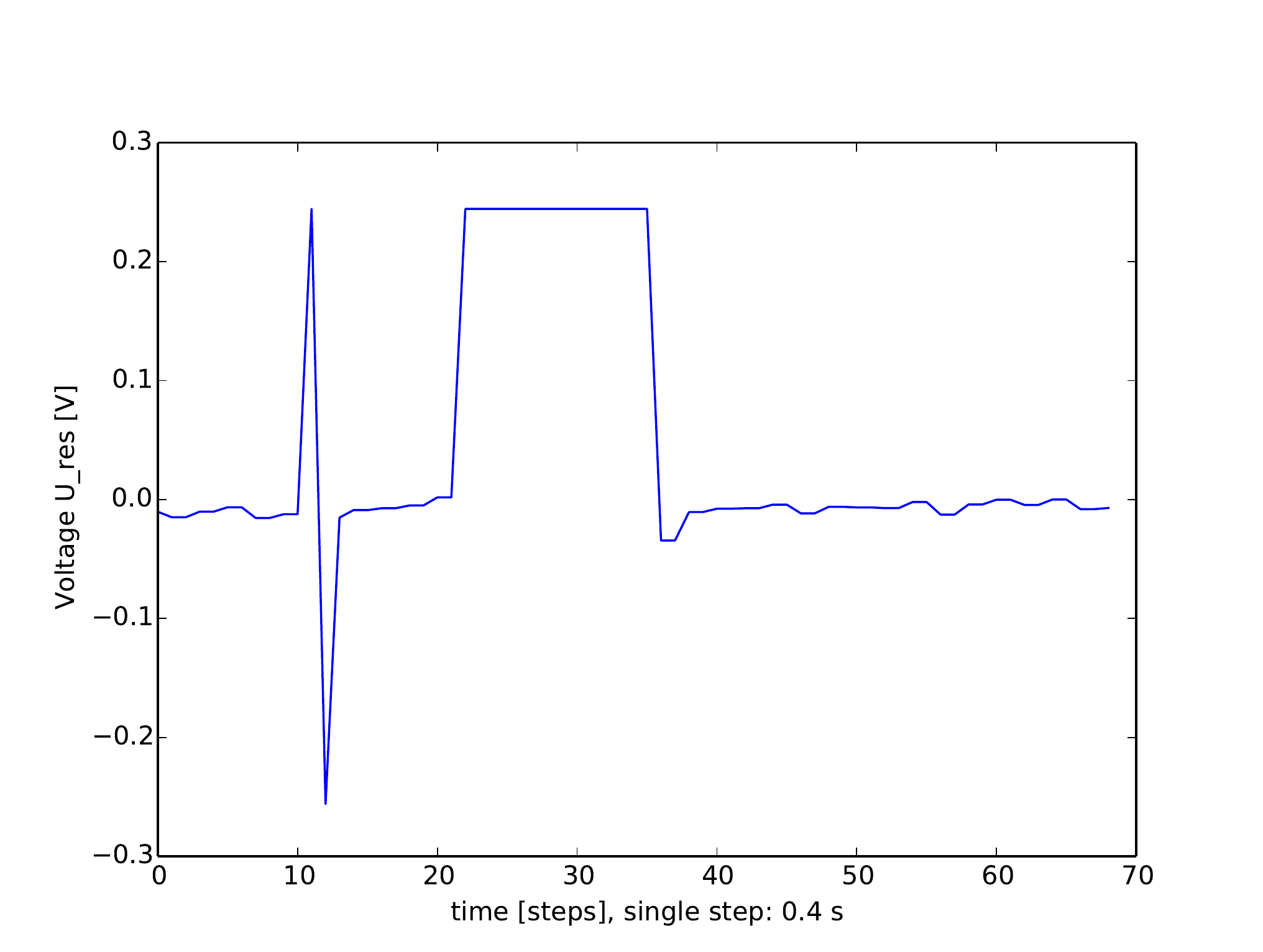}
\end{subfigure}
\begin{subfigure}{.4\textwidth}
  \centering
  \includegraphics[width=.9\linewidth]{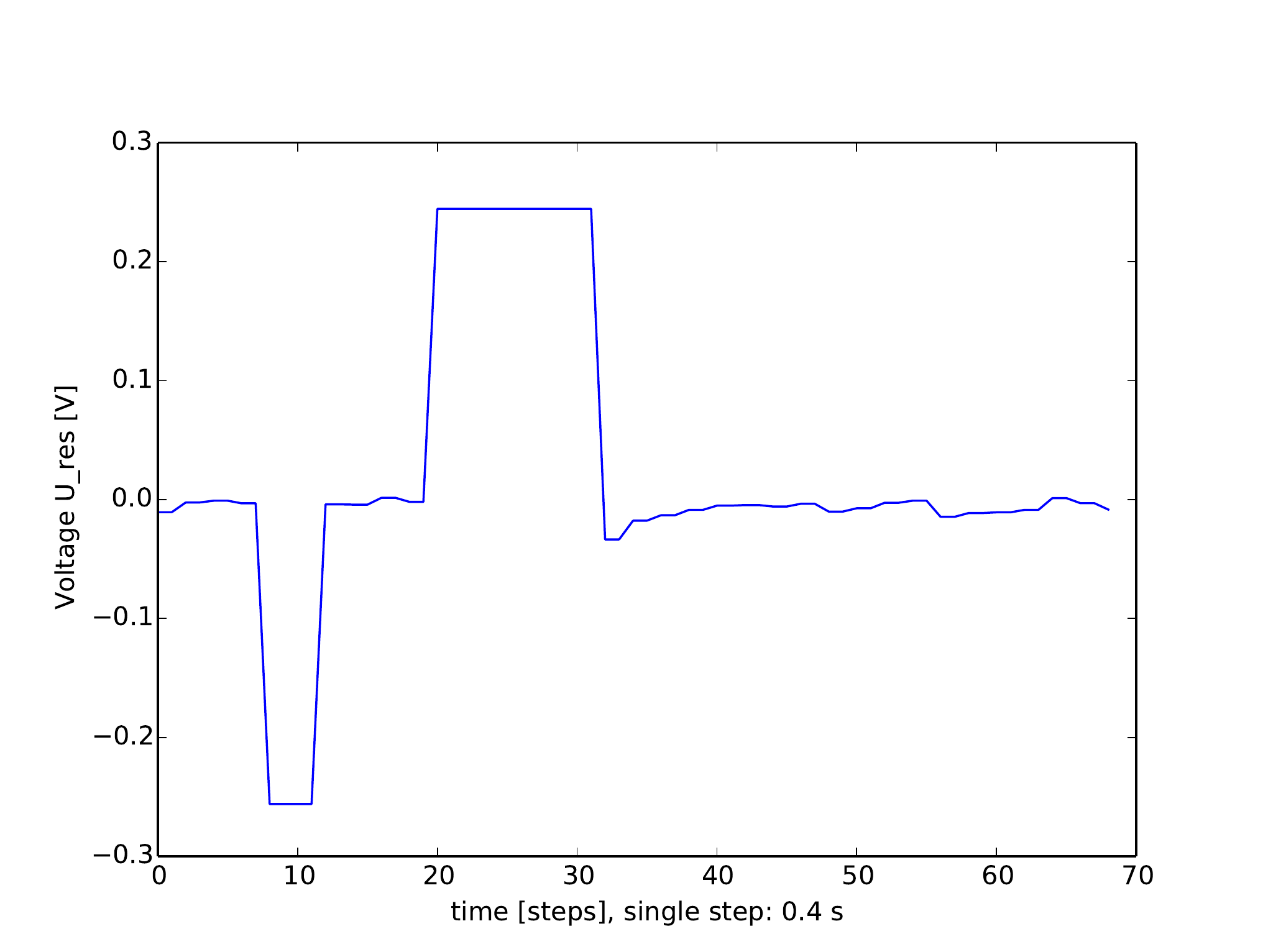}
\end{subfigure}%
\begin{subfigure}{.4\textwidth}
  \centering
  \includegraphics[width=.9\linewidth]{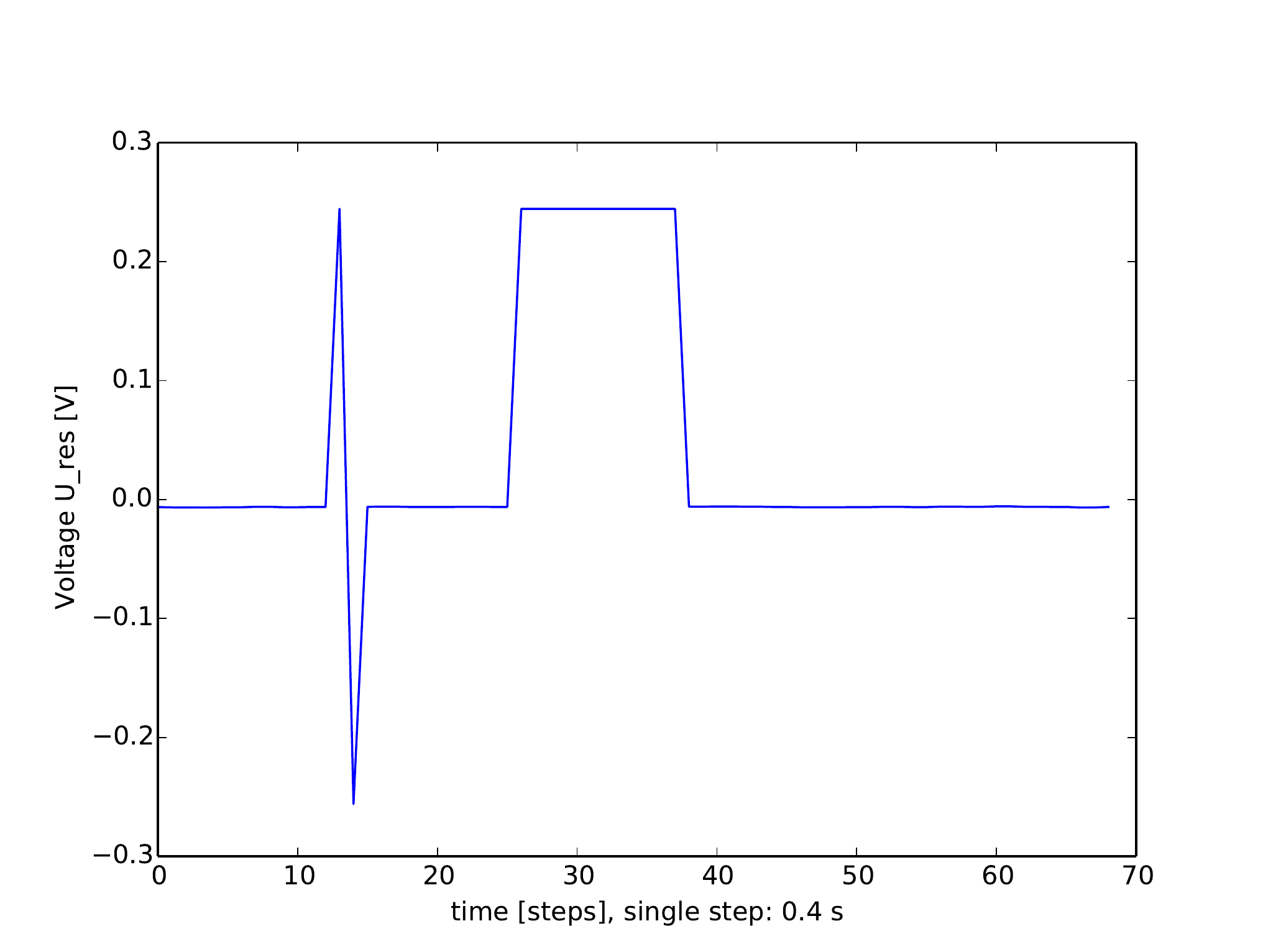}
\end{subfigure}
\caption{The selected sample anomalies of $600\,A$ magnets extracted from the LS database.}
\label{fig:samples_quenches}
\end{figure*}

There are various kinds of magnets located in the LHC tunnel and they generate different number of quench events (Tab.~\ref{tab:LHCcircuits}, Fig. \ref{fig:samples_quenches}). It is beneficial to choose a group of magnets for which the largest possible number of quenches was recorded. The longest history of quenches was provided for \SI{600}{\ampere} magnets in LS database. Therefore, we decided to focus our initial research on \SI{600}{\ampere} magnets. Unfortunately, the \SI{600}{\ampere} magnets data stored in a database is very large i.e. an order of several gigabytes. However, as it was mentioned before, the activity record of superconducting magnets during operational time of the LHC is composed mostly of sections of normal operation and only sometimes quench events happen. Furthermore, the logging database does not enable automated quench periods extraction, despite having many useful features for a data preprocessing and information extraction.

\begin{figure}
\hspace{-0.7cm}
\includegraphics[width=1.2\hsize]{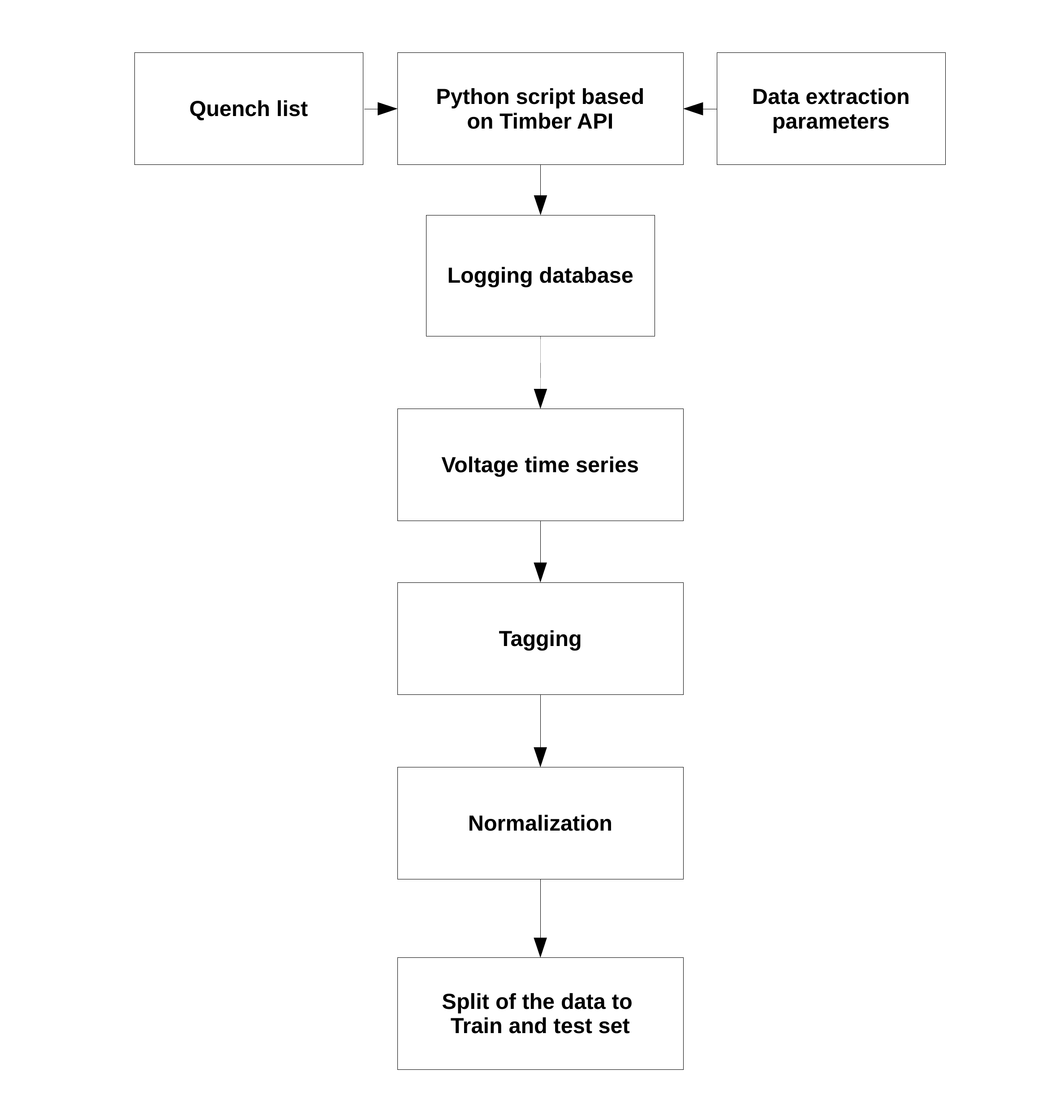}
\caption{The procedure for extraction of voltage time series with anomalies from the LS database.}
\label{fig:data_extraction}
\end{figure}

It would be a tedious work to manually extract all the quenches. Therefore, a quench extraction application (presented in Fig. \ref{fig:data_extraction}) was developed, which automates the process of fetching the voltage time series from the LS database. It is composed of a set of Python scripts which generate appropriately prepared queries to the LS database. The queries are built based on the quench list \citep{LHC_quench_db} and data extraction parameters configuration files. Once the data has been fetched from the LS database it is normalized to the range from \num{0} to \num{1} and split into training and testing set: \SI{70}{\percent} of the data is used for training and \SI{30}{\percent} for testing.

Different lengths of time window frame before and after the quench events were considered. Ultimately, we chose in our view a reasonable trade-off between the amount of data and their representativeness for the model i.e. \num{24} hour long time window before a quench event. We extracted days on which quenches occurred between the years 2008 and 2016, which amounted to \num{425} in total for \SI{600}{\ampere} magnets (Tab.~\ref{tab:LHCcircuits}).

A training of deep learning models takes long time even when fast GPUs are employed for the calculations. Therefore, it is important what kind of and how large data sets are used for training and testing the models. Furthermore, it is important to preliminary adjust hyper-parameters of the model using relatively small data set when a single iteration time is short. Thereafter, tiny updates are done using the largest dataset, when each training routine of the network consumes substantial amount of time. Thus, we have created three different data sets: small, medium and the large ones as presented in Tab.~\ref{tab:data_sets}.

\begin{table}
\caption{The data sets used for training and testing the model.}
\label{tab:data_sets}
\centering
\begin{tabular}{lc}
\toprule
Data set & size [\si{\mega\byte}] \\ 
\midrule
small & \num{22}  \\
medium & \num{111} \\
large & \num{5000}  \\
\bottomrule
\end{tabular}
\end{table}

\subsection{Results and analysis}

\begin{figure}
\centering
\includegraphics[width=0.5\hsize]{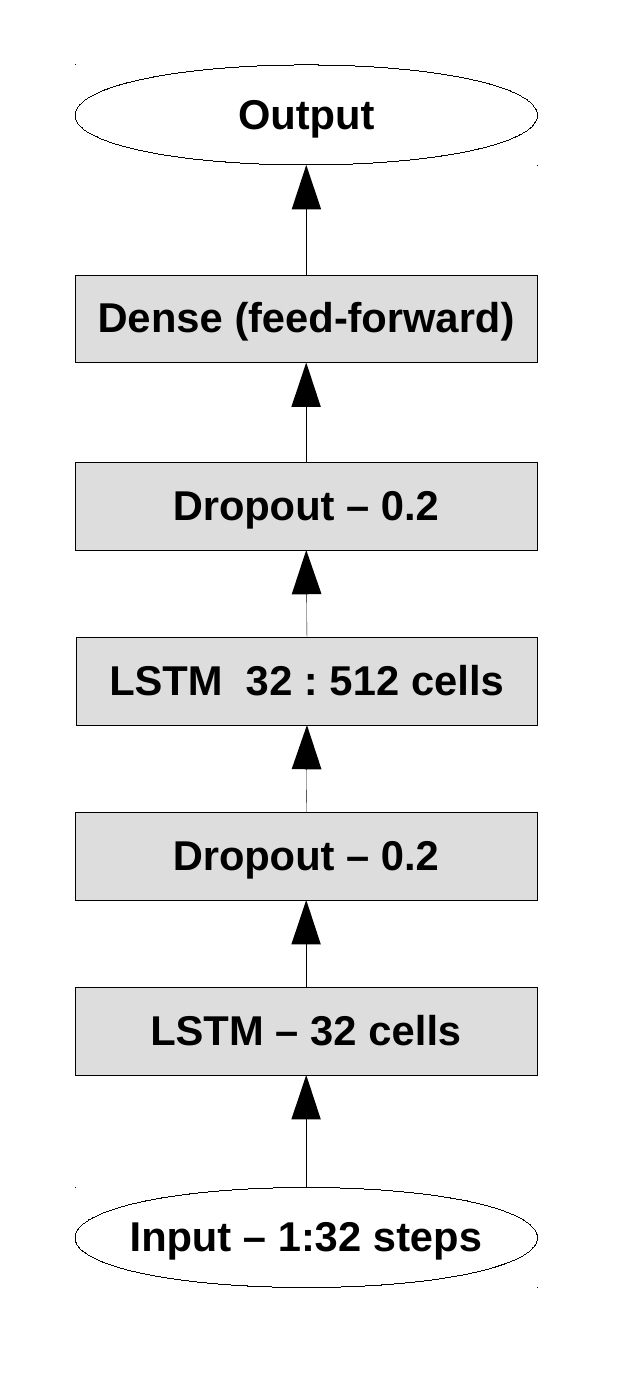}
\caption{The LSTM-based network used for the experiments.}
\label{fig:lstm_architecture_experiments}
\end{figure}

The designed LSTM model was examined with a respect to its ability of anticipating few voltage values forward. The core architecture of the module used for the experiments is presented in Fig. \ref{fig:lstm_architecture_experiments}, but we trained and tested various models with wide range of parameters such as number of neurons, layers and inputs. The primary goal was to find a core set of the parameters which enabled modeling of the $U_{res}$ voltage data. 

The core architecture of the network module is composed of seven layers all together: an input layer, four LSTM hidden layers, one feed-forward hidden layer and an output layer. It is worth noting that dropout operations are also classified as separate layers. Every second layer among layers of the LSTM type has a dropout with a value of \SI{20}{\percent}. Furthermore, a number of LSTM cells in the middle layer were changed from \num{32} to \num{128} and then to \num{512} in order to examine the performance of the model as a function of the number of neurons. The module was implemented in Keras \citep{keras} with the Theano backend \citep{theano}.

\begin{figure*}
\hspace{-1.0cm}
\begin{subfigure}{.6\textwidth}
  \centering
  \includegraphics[width=0.9\linewidth]{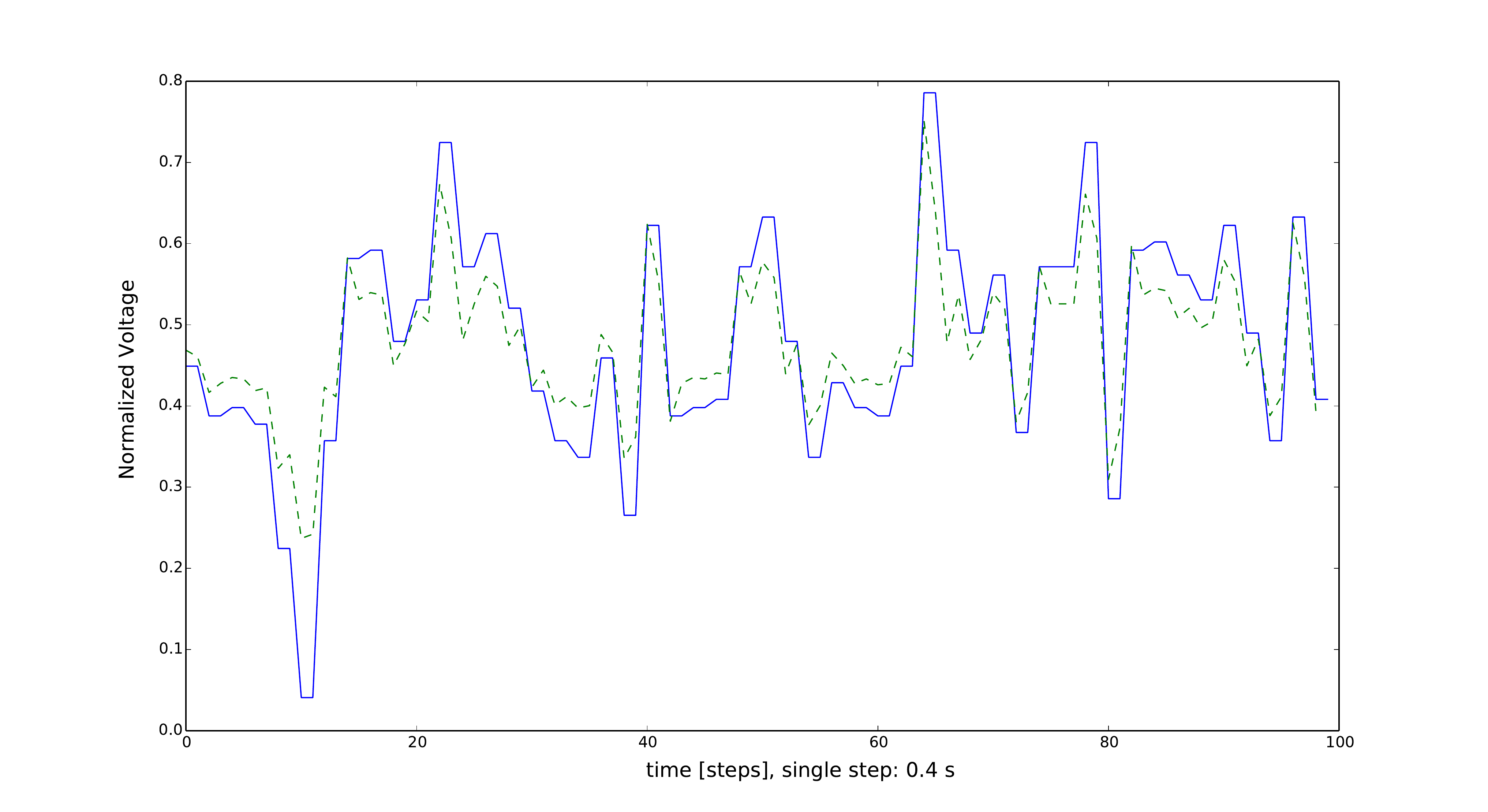}
  \caption{One step ahead}
  \label{fig:prediction_back-2_future-1}
\end{subfigure}%
\begin{subfigure}{.6\textwidth}
  \centering
  \hspace{-1.0cm}
  \includegraphics[width=0.9\linewidth]{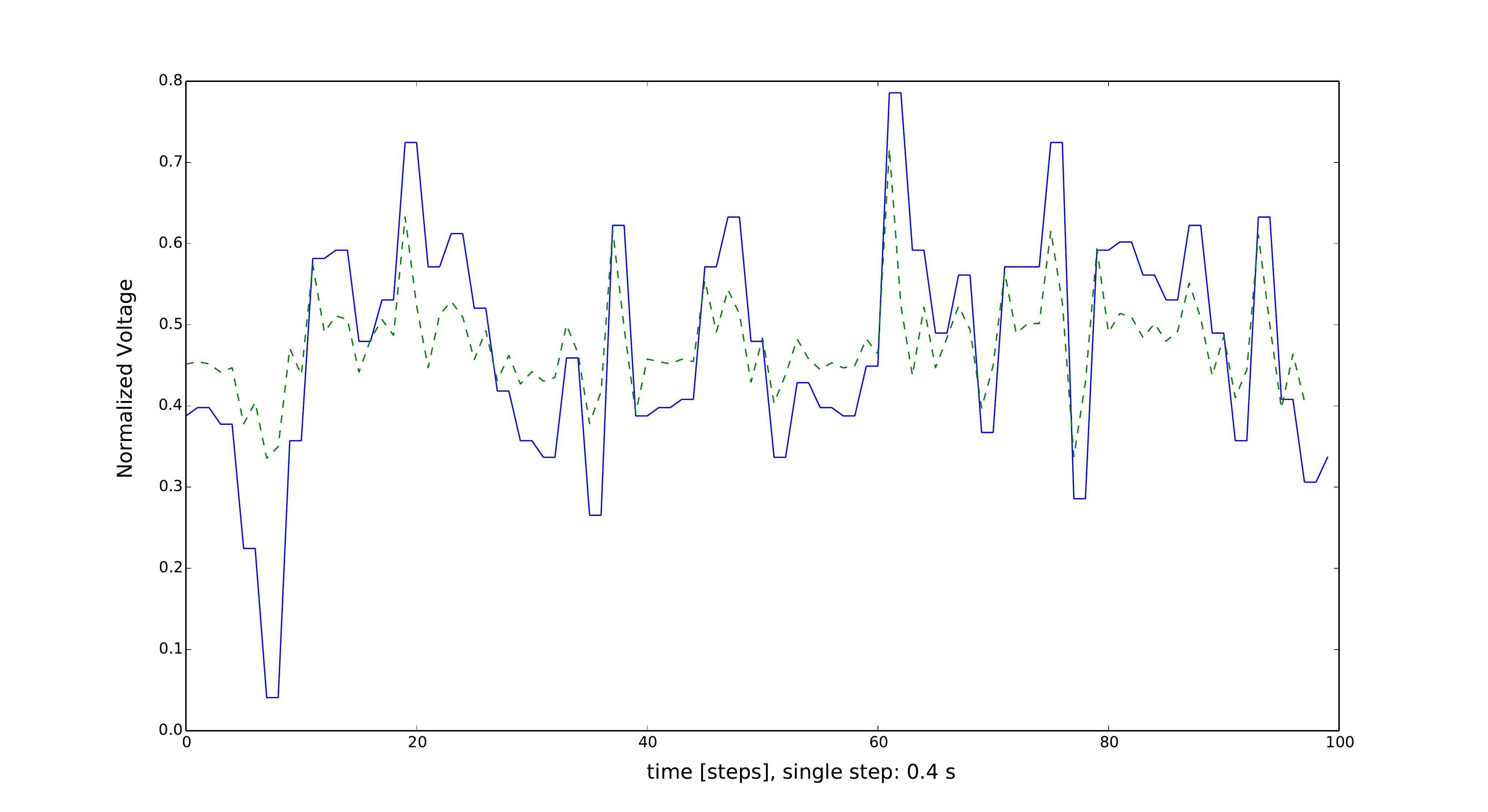}
  \caption{Two steps ahead}
  \label{fig:prediction_back-2_future-2}
\end{subfigure}
\caption{Two examples of prediction for one and two steps ahead in time. Predicted signal is plotted in a green broken line.}
\label{fig:prediction_back_future}
\end{figure*}

Fig. \ref{fig:prediction_back_future} shows both real voltage signal and its prediction. Visual similarity analysis is neither efficient nor recommended for validation of regression models therefore authors have decided to use more reliable measures such as Root Mean Square Error (RMSE) and Mean Percentage Error (MPE).

The measures are given by the following equations:

\begin{equation}
\mathit{RMSE} = {\sqrt {\frac{1} {N}{\sum\limits_{t = 1}^N {({y} ^{(t)} - \hat{y} ^{(t)}} })^{2} }} 
  \label{eq:rmse}
\end{equation}

\begin{equation}
\mathit{MPE} = {\frac{100 \%} {N}{\sum\limits_{t = 1}^N \frac{{{y} ^{(t)} - \hat{y} ^{(t)}}}{{y} ^{(t)}} } }
  \label{eq:mpe}
\end{equation}

\noindent 
where: ${y} ^{(t)}$ and $\hat{y} ^{(t)}$ is a voltage time series and its predicted counterpart, respectively. Both equations (\ref{eq:rmse}) and (\ref{eq:mpe}) are calculated for $N$ data points which in turn depend on the size of data set that is used to train and test the model.

\begin{figure*}
\centering
\includegraphics[width=1\hsize]{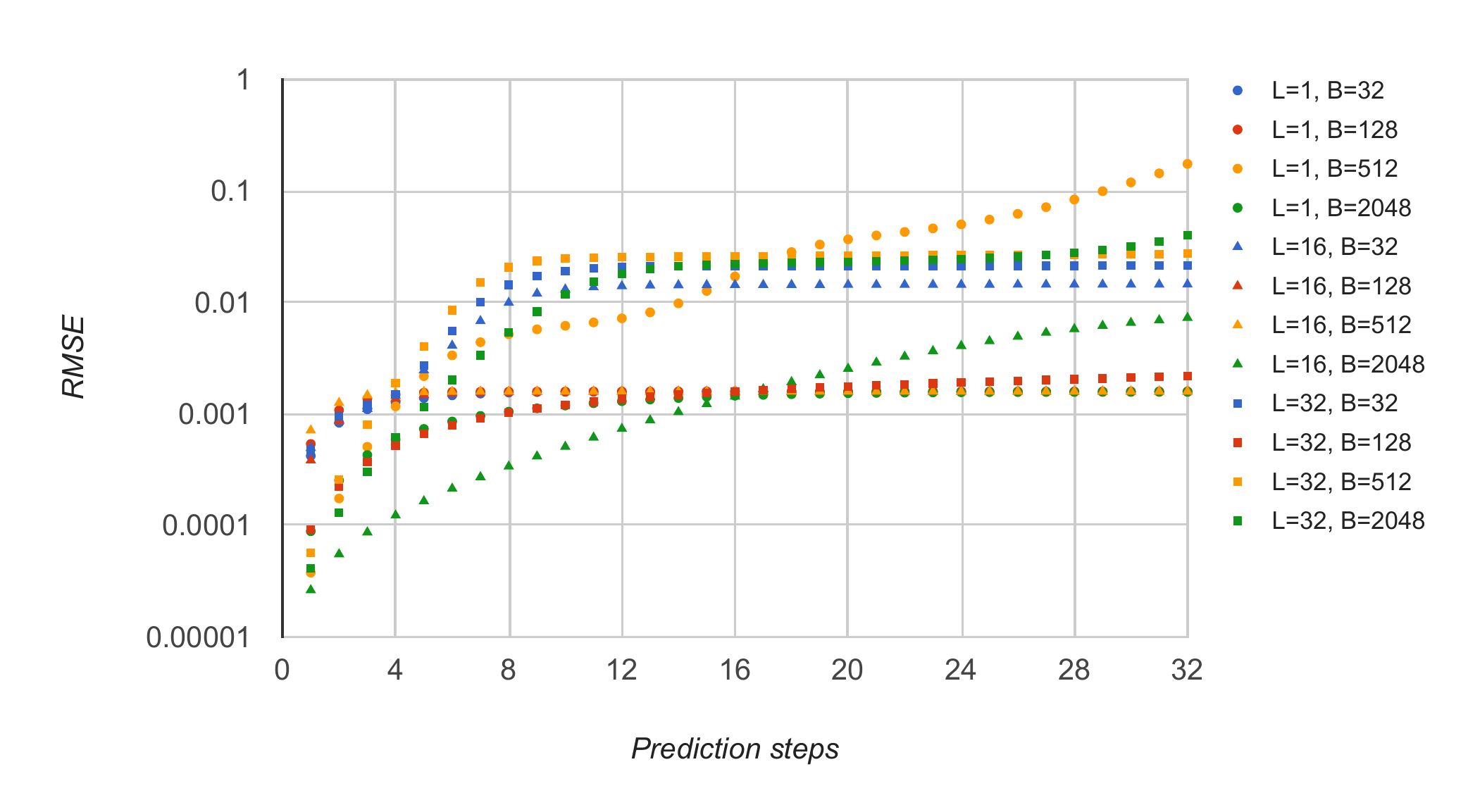}
\caption{The value of RMSE as a function of prediction steps for different batch size $B$ and number of previous time steps $L$ values with \num{32} neurons in the middle LSTM layer.}
\label{fig:scatter-plot_32-neurons}
\end{figure*}

\begin{figure*}
\centering
\includegraphics[width=1\hsize]{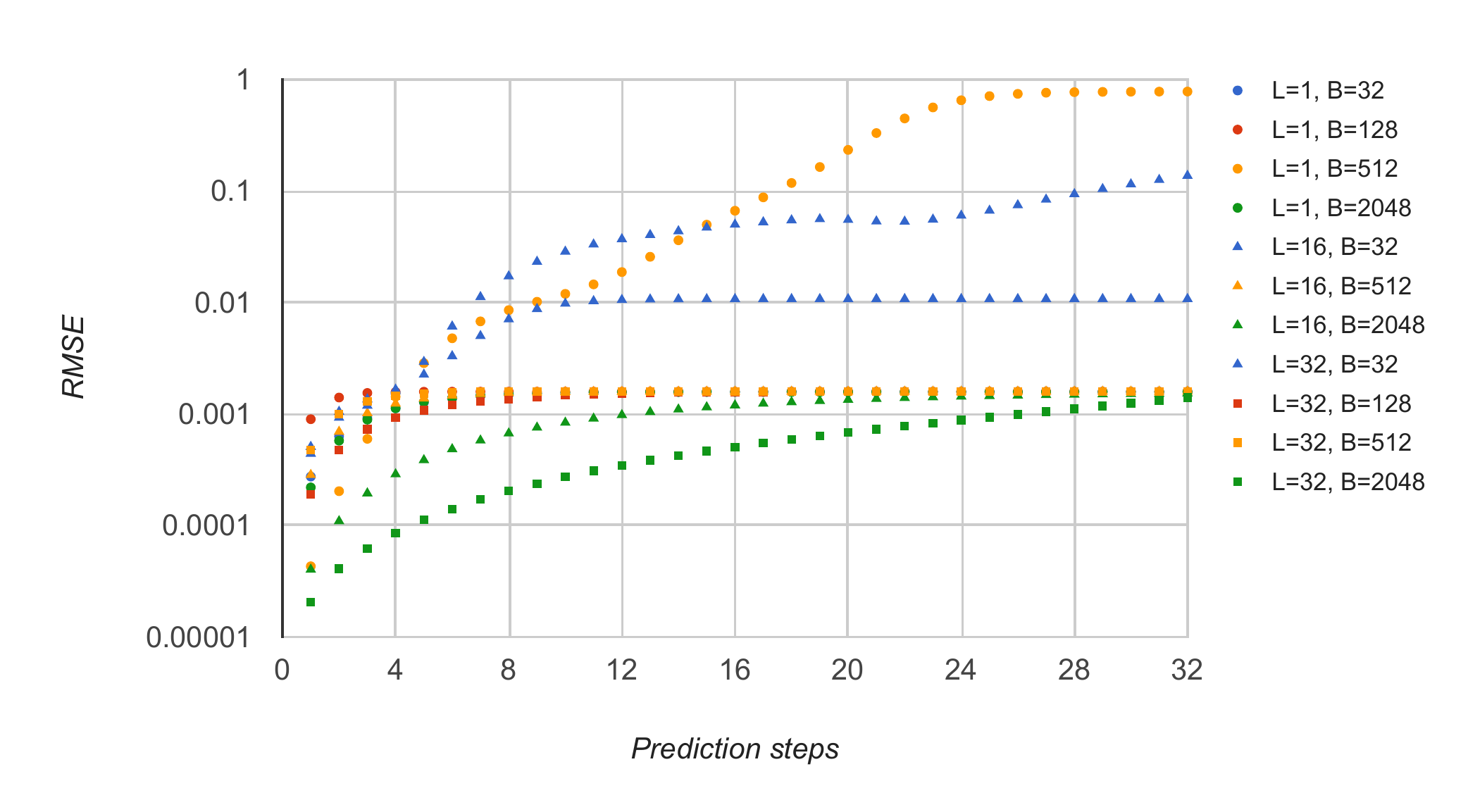}
\caption{The value of RMSE as a function of prediction steps for different batch size $B$ and number of previous time steps $L$ values with \num{128} neurons in the middle LSTM layer.}
\label{fig:scatter-plot_128-neurons}
\end{figure*}

\begin{figure*}
\centering
\includegraphics[width=1\hsize]{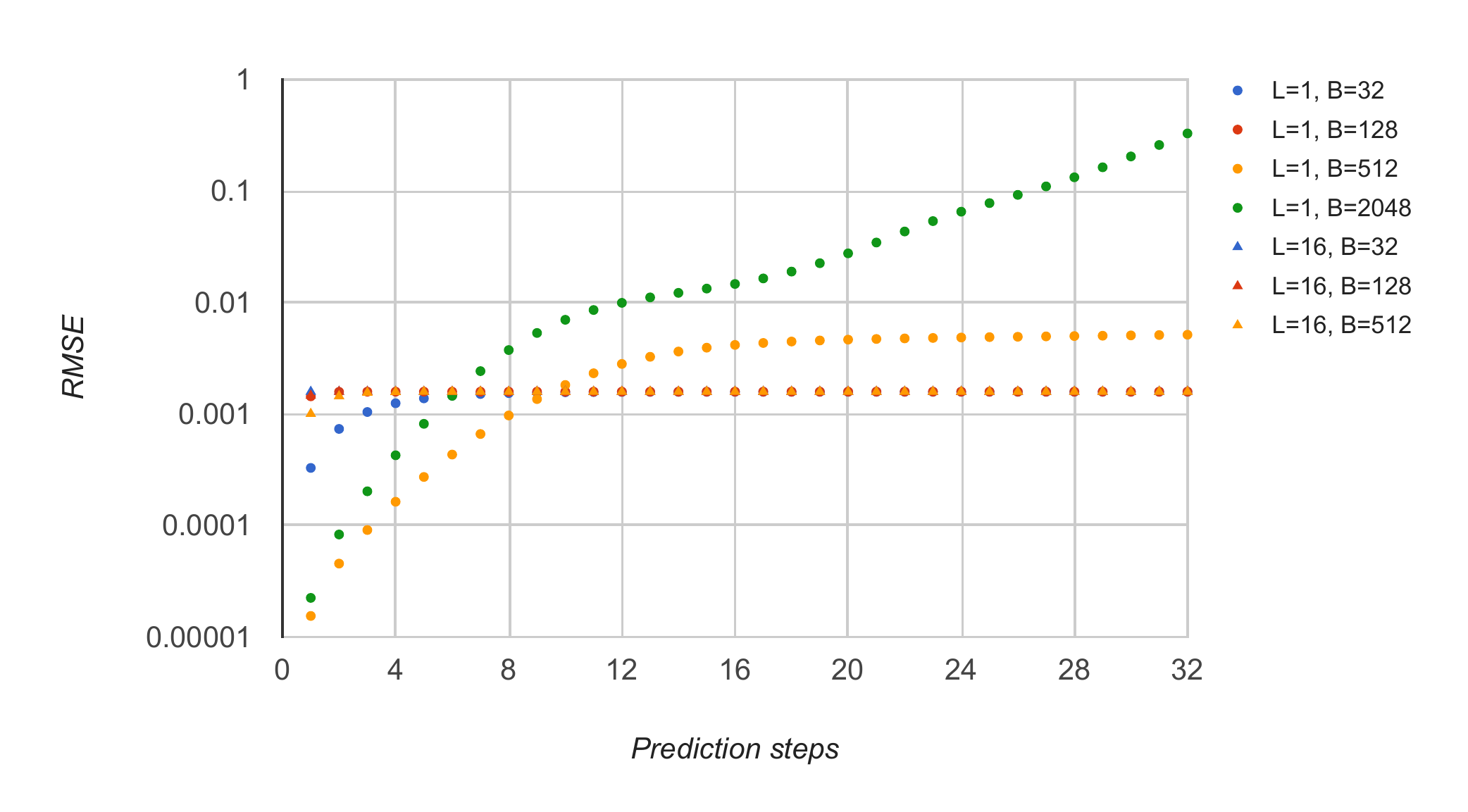}
\caption{The value of RMSE as a function of prediction steps for different batch size $B$ and number of previous time steps $L$ values with \num{512} neurons in the middle LSTM layer.}
\label{fig:scatter-plot_512-neurons}
\end{figure*}

Fig.~\ref{fig:scatter-plot_32-neurons} -- \ref{fig:scatter-plot_512-neurons} present the prediction results in terms of RMSE for \num{32} future steps. The model was trained using the medium data corpus. Two quantities were used as parameters:
\begin{description}
\item[$L$] - a number of previous time steps to use as input variables to predict the next time period,
\item[$B$] - a size of training batch.
\end{description}
\noindent
The experiments were conducted for three different $L$ values: \num{1}, \num{16} and \num{32} which also affected the model input size. The more steps back in time are taken into account in model training and testing processes the wider input should be used. The size of the model input is equal to a number of steps back $L$ in time which are taken for building the LSTM model. Furthermore, four different batch sizes $B$ values were tested: \num{32}, \num{128}, \num{512} and \num{2048}. The batch size $B$ has two-fold effect on the performance of the model. On the one hand it affects a range of the voltage series which is processed by the model. On the other hand, the larger batches are computed faster on GPUs because matrix calculation optimization measures may be applied.

According to Fig.~\ref{fig:scatter-plot_32-neurons}, the best result (in terms of mean RMSE) in the experiment with \num{32} neurons used in the middle LSTM layer was obtained for $L=\num{16}$ and batch size $B=\num{2048}$. In the case of the experiments with \num{128} neurons, the parameters combination of $L=\num{32}$ and batch size $B=\num{2048}$ resulted in the best result in terms of RMSE as it is presented in Fig.~\ref{fig:scatter-plot_128-neurons}.

\begin{figure*}
\centering
\includegraphics[width=.7\hsize]{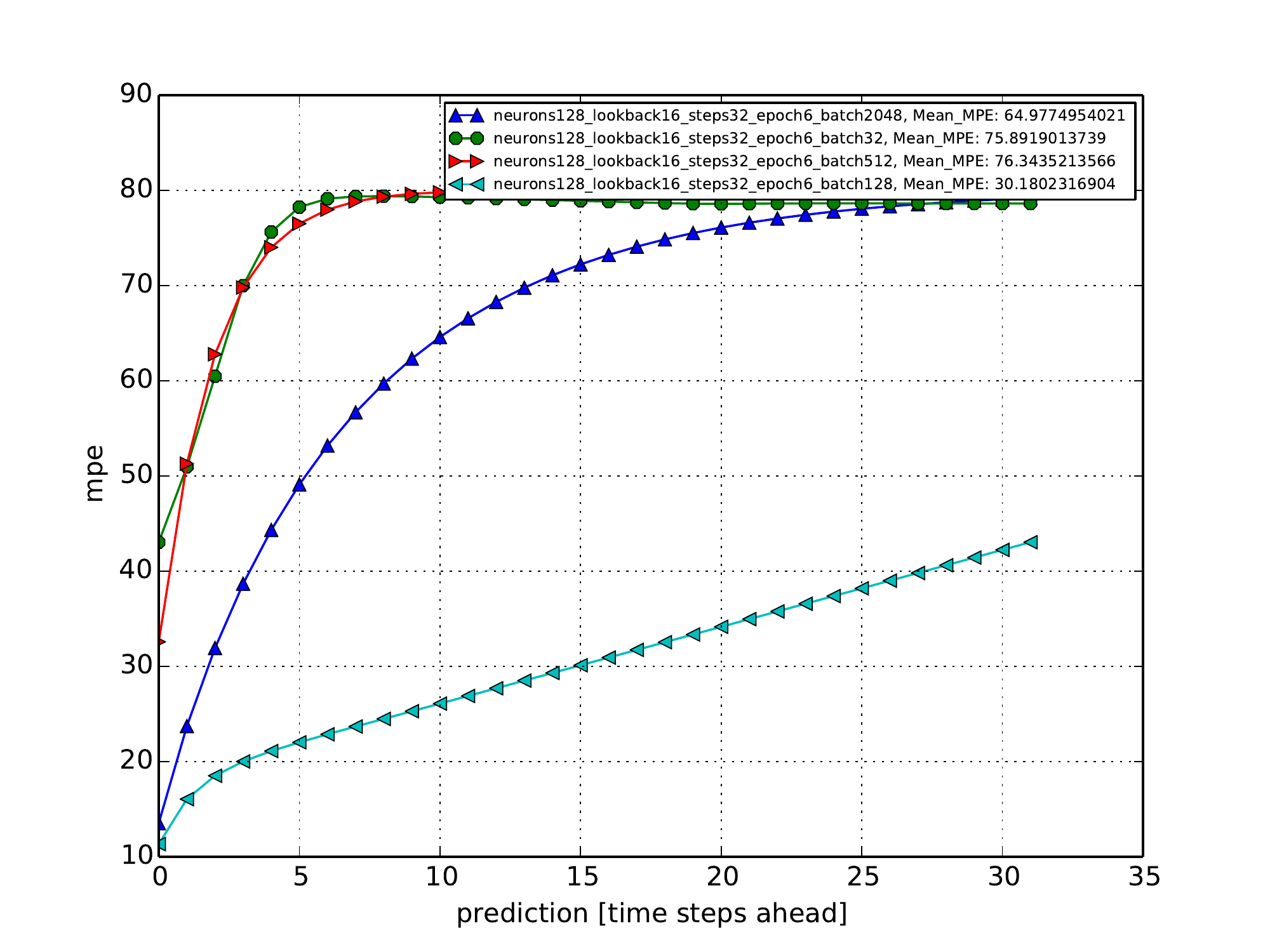}
\caption{The example of the MPE plot as a function of prediction steps for different batch size $B$ and number of previous time steps $L$ values.}
\label{fig:figure_mpe_sample}
\end{figure*}

\begin{figure*}
\centering
\includegraphics[width=.7\hsize]{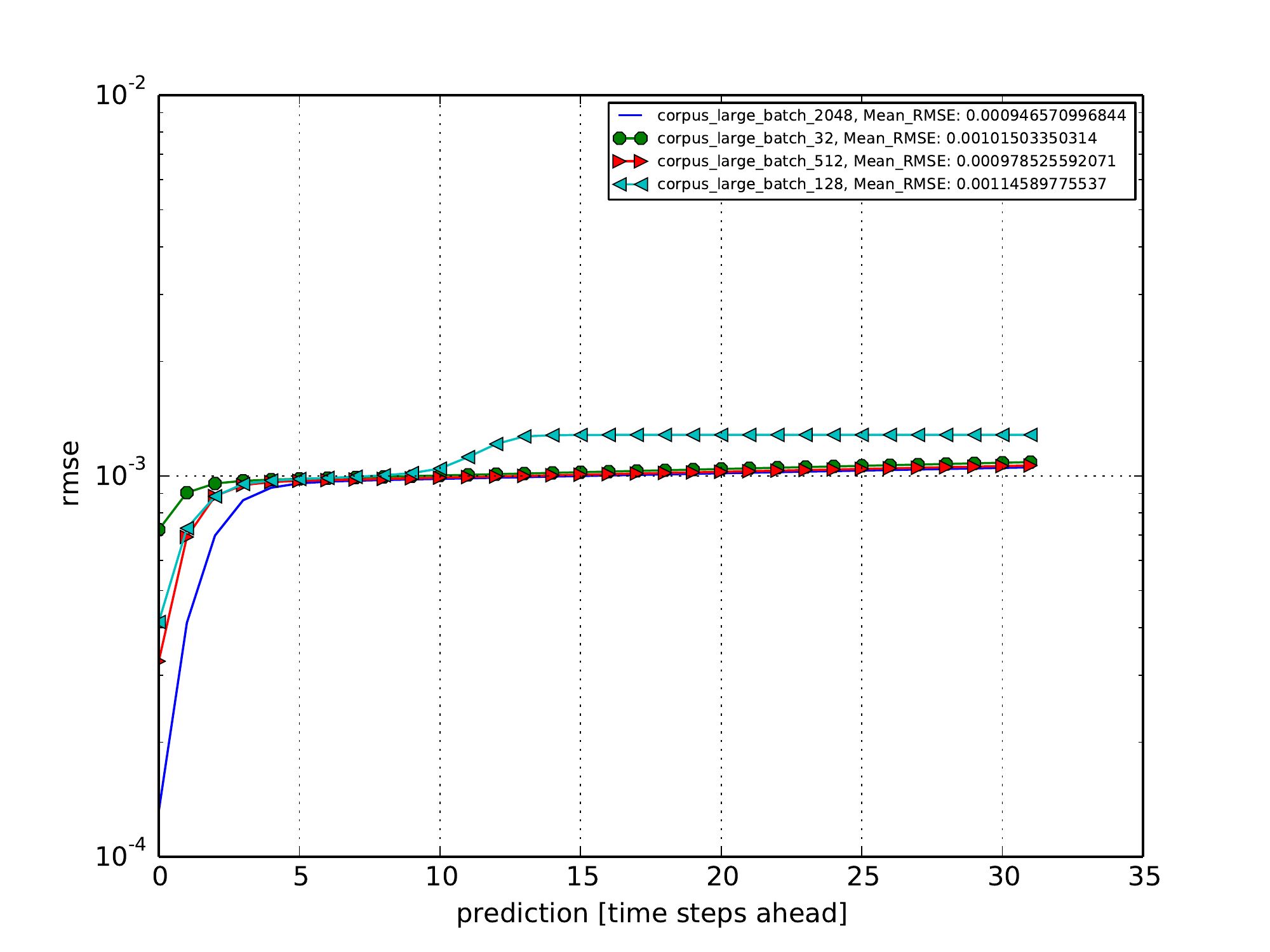}
\caption{The example of the RMSE plot as a function of prediction steps for the large corpus (Tab.~\ref{tab:data_sets}).}
\label{fig:figure_rmse_large_corpus}
\end{figure*}

The MPE values, according to equation (\ref{eq:mpe}), were computed for a selected quench fragments in parallel to the RMSE calculation of the whole voltage time series. The results of the MPE calculation generally follow the RMSE trend, therefore we decided not to include them. However, a sample MPE plot is presented in Fig.~\ref{fig:figure_mpe_sample}. It is worth noting that a prediction quality is lower for more steps ahead. This is due to the fact the a prediction of $t_{n+1}$ time step is based on a previous $t_{n}$ step.

\begin{table}
\caption{The best results obtained for the medium corpus (Tab.~\ref{tab:data_sets}).}
\label{tab:best_results_medium_corpus}
\centering
\begin{tabular}{cccc}
\toprule
LSTM cells & $L$ & $B$ & mean RMSE\\ 
\midrule
\num{128} & \num{16} & \num{2048} & \num{0.00104} \\
\num{32} & \num{1} & \num{2048} & \num{0.00125} \\
\num{128} & \num{32} & \num{128} & \num{0.00140} \\
\num{32} & \num{1} & \num{32} & \num{0.00148} \\
\bottomrule
\end{tabular}
\end{table}

Tab.~\ref{tab:best_results_medium_corpus} presents the results in terms of the mean value of the RMSE. The results were obtained by averaging the RMSE error over all the steps in the future. It is worth emphasizing that the lowest RMSE error is achieved for predictions with a small number of forward steps. The more steps to the future are predicted, the worse results are obtained which is reflected in the rising RMSE value. Therefore, the results provided by Tab.~\ref{tab:best_results_medium_corpus} should be considered as an approximate performance of the LSTM model. Nevertheless, it is noticeable that the best two results are achieved for batch size of \num{2048}.

\begin{table}
\caption{The parameters of the LSTM network used to the experiments.}
\label{tab:test_setup_parameters}
\centering
\begin{tabular}{lc}
\toprule
Parameter & Value \\ 
\midrule
Number of layers & \num{5} \\
Number of epochs & \num{6} \\
Total number of the network parameters & \num{21025} \\
Dropout & \num{0.2} \\
Max. number of steps ahead & \num{32} \\
\bottomrule
\end{tabular}
\end{table}

All the tests presented in this section were performed on Intel(R) Core(TM) i7-4790 CPU @ \SI{3.60}{\giga\hertz} with \SI{32}{\giga\byte} DDR3 \SI{1600}{\mega\hertz} memory. The processing time was long - it took over two weeks to compute the presented results. The biggest contribution was the training time of the LSTM model, which was significantly higher for the architectures with more neurons. This was the main reason why the model was trained over only six epochs as presented in Tab.~\ref{tab:test_setup_parameters}. 

A test for the large corpus (Tab.~\ref{tab:data_sets}) of \SI{5}{\giga\byte} was also conducted and an example of the results is presented in Fig.~\ref{fig:figure_rmse_large_corpus}. Because of the long computation time we were not able to conduct the same set of experiments as we did for the medium size corpus. Nevertheless, the results we managed to gather show that the RMSE value converges to a value on the level of \num{0.001}. This was expected since much more data was used for the experiment with the large corpus. 

\section{Discussion}
\label{section:discussion}

\subsection{Solution applicability}

Safety of the systems used in LHC is of a high importance. As a result, any neural-network-based application is best thought of as an addition to or enhancement of the highly dependable current system.

There is also a broad range of possible applications of RNN-based solutions in the other CERN control systems, such as cryogenics, vacuum, machine protection and power converters \citep{marin2017fault, tsan2016cern}. Examples include:

\begin{itemize}
\item anomaly detection on beam screen,
\item faulty cryogenics valve detection, and
\item Future Circular Collider (FCC) Reliability, Availability, Maintainability and Safety (RAMS) studies.
\end{itemize}

\subsection{Anomalies classification}

It is worth emphasizing that in order to classify anomalies it is essential to map regression results to classification task. In other words, it would be necessary to express RMSE in terms of F1 score. Unfortunately, this would require well defined threshold of RMSE value, which would discriminate positive and negative classification results \citep{nanduri2016anomaly}. At the current stage of the research, such a threshold value has not been determined and requires further investigation. It may also be possible to adopt different approaches to the conversion to a classification task, especially ones not requiring data of anomalous behavior during model training \citep{malhotra2015long,bontemps2016collective,o2016recurrent}. This is a subject of the ongoing research.

\subsection{System reaction time}

The authors are currently implementing RNN LSTM module on FPGA to be used for anomaly detection. However, it is worth noting that networks of the similar size as the one which is to be used for the system described in the paper were already implemented and described in the following papers \citep{chang2015recurrent, han2017ese, lee2016fpga}. The performance of the LSTM module implemented on FPGA strictly depends on three main factors:
\begin{itemize}
    \item memory footprint of the network,
    \item localization of the network weights (external or internal memory),
    \item degree to which the network is compressed.
\end{itemize}

It is worth emphasizing that the network weights (coefficients) are used for every single iteration of the computations. Therefore keeping them in the internal memory of the processing unit (FPGA) is highly beneficial for the performance of the module.

\begin{table*}
\caption{Performance of various approaches to LSTM hardware implementation (data from \citep{chang2015recurrent, han2017ese, lee2016fpga}).}
\label{tab:hardware_lstm}
\centering
\begin{tabular}{lll}
\toprule
Setup & Platform & \makecell[l]{Computation\\time [\si{\micro\second}]} \\
\midrule
\makecell[l]{2 layers (\num{128} cells),\\32/16 bit} & \makecell[l]{Xilinx Zynq 7020 (\SI{142}{\mega\hertz}),\\external memory - DDR3} & \num{\sim 932} \\
\addlinespace[0.5em]
\makecell[l]{compressed LSTM (20x),\\\num{1024} cells} & \makecell[l]{Xilinx XCKU060 Kintex (\SI{200}{\mega\hertz}),\\ external memory - DDR3} & \num{82.7} \\
\addlinespace[0.5em]
\makecell[l]{2 layers (\num{30}, \num{256} cells),\\6-bit quantization} & \makecell[l]{Xilinx Zynq XC7Z045 (\SI{100}{\mega\hertz})\\\SI{2.18}{\mega\byte} on-chip memory max,\\all in the internal memory} & \num{15.96} \\
\bottomrule
\end{tabular}
\end{table*}

Tab.~\ref{tab:hardware_lstm} shows results gathered from three papers which adopt various approaches to illustrate discrepancies in performance across them.

The first approach \citep{chang2015recurrent} shows uncompressed network (with high data representation precision) and the weights are stored in the external memory. It takes roughly \SI{1}{\milli\second} to process a single LSTM iteration.

The second approach \citep{han2017ese} takes advantage of weights compression which affects significantly the performance of the module. The weights were reduced to \num{12} bits which resulted in a drop of an amount data which is fetched from the memory for the computations in each iteration of the algorithm. Consequently, the module is capable of achieving \SI{\sim 83}{\micro\second} for the single LSTM iteration processing time.

The third approach \citep{lee2016fpga} adopts both coefficients compression and optimal weights localization. All the coefficients are kept in the internal BRAM memory. This is possible due to the high compression ratio of the weights. Huge gain in the performance may be noticed compared to the two previously presented approaches (single iteration takes \SI{\sim 16}{\micro\second}).

It is worth noting that the further performance improvement is possible by moving coefficients from BRAM memories to the distributed ones.

\section{Conclusions and future work}
\label{section:conclusions}

The data, including quench events as sample anomalies, acquired from the logging database were used to verify LSTM recurrent neural networks ability to model voltage time series of LHC superconducting magnets. It has been proved that LSTM-based setup performs well, with RMSE value approaching \num{0.001} for the largest dataset used. As it was expected, prediction results for more steps ahead are inferior to the short time prediction in terms of accuracy expressed in RMSE.

As a future work, the Post Mortem data of much bigger resolution will be used for model training and the algorithm for anomalies detection and/or prediction will be developed and verified. Authors are also going to implement the prediction stage of the algorithm in FPGA or ASIC to evaluate its performance in real-time applications. 

\vspace{0.5em}

\vspace*{1.5em}

\bibliography{mybibfile}

\end{document}